\documentclass[12pt,letter]{article}
\usepackage{amsfonts,amssymb,amsmath}
\numberwithin{equation}{section}
\begin{document}
\newcommand{\tr}{\operatorname{tr}}
\newcommand{\ds}{\displaystyle}
\newcommand{\e}{\operatorname{e}}
\newcommand{\mustbe}{\stackrel{!}{=}}
\newcommand{\rt}{(r,\vartheta)}
\newcommand{\rts}{(r',\vartheta')}

\newcommand{\sdot}{\,{\scriptscriptstyle{}^{\bullet}}\,}

\newcommand{\trip}{{}^3S_1}
\newcommand{\sing}{{}^1S_0}
\newcommand{\A}{\mathcal{A}}

\newcommand{\ap}{{}^{(a)}\varphi_{\pm}(\vec{r})}
\newcommand{\app}{{}^{(a)}\varphi_{+}(\vec{r})}
\newcommand{\apm}{{}^{(a)}\varphi_{-}(\vec{r})}

\newcommand{\sapm}{{}^{(a)} \varphi_{-}^{\dagger}(\vec{r})}
\newcommand{\sapp}{{}^{(a)} \varphi_{+}^{\dagger}(\vec{r})}

\newcommand{\ipp}{{}^{(1)}\varphi_{+}(\vec{r})}
\newcommand{\ipm}{{}^{(1)}\varphi_{-}(\vec{r})}
\newcommand{\iipp}{{}^{(2)}\varphi_{+}(\vec{r})}
\newcommand{\iipm}{{}^{(2)}\varphi_{-}(\vec{r})}
\newcommand{\sPh}{{\stackrel{*}{\Phi}}(\vec{r}_1,\vec{r}_2)}
\newcommand{\sph}{{\stackrel{*}{\varphi}}}
\newcommand{\sphi}{{\stackrel{*}{\varphi}_1}(\vec{r}_1)}
\newcommand{\sphii}{{\stackrel{*}{\varphi}_2}(\vec{r}_2)}
\newcommand{\sphis}{{\stackrel{*}{\varphi}_1}(\vec{r}\,')}
\newcommand{\sphiis}{{\stackrel{*}{\varphi}_2}(\vec{r}\,')}

\newcommand{\aRpm}{{}^{(a)\!}\tilde{R}_\pm}
\newcommand{\aRp}{{}^{(a)\!}\tilde{R}_+}
\newcommand{\aRm}{{}^{(a)\!}\tilde{R}_-}
\newcommand{\iRp}{{}^{(1)\!}\tilde{R}_+}
\newcommand{\iRm}{{}^{(1)\!}\tilde{R}_-}
\newcommand{\iiRp}{{}^{(2)\!}\tilde{R}_+}
\newcommand{\iiRm}{{}^{(2)\!}\tilde{R}_-}

\newcommand{\aSpm}{{}^{(a)\!}\tilde{S}_\pm}
\newcommand{\aSp}{{}^{(a)\!}\tilde{S}_+}
\newcommand{\aSm}{{}^{(a)\!}\tilde{S}_-}
\newcommand{\iSp}{{}^{(1)\!}\tilde{S}_+}
\newcommand{\iSm}{{}^{(1)\!}\tilde{S}_-}
\newcommand{\iiSp}{{}^{(2)\!}\tilde{S}_+}
\newcommand{\iiSm}{{}^{(2)\!}\tilde{S}_-}

\newcommand{\ikn}{{}^{(1)}\!k_0(\vec{r})}
\newcommand{\ikp}{{}^{(1)}\!k_\phi(\vec{r})}
\newcommand{\iikp}{{}^{(2)}\!k_\phi(\vec{r})}

\newcommand{\iikn}{{}^{(2)}\!k_0(\vec{r})}
\newcommand{\ikns}{{}^{(1)}\!k_0(\vec{r}\,')}
\newcommand{\iikns}{{}^{(2)}\!k_0(\vec{r}\,')}

\newcommand{\itkn}{{}^{(1)}\!\tilde{k}_0}
\newcommand{\iitkn}{{}^{(2)}\!\tilde{k}_0}

\newcommand{\itkp}{{}^{(1)}\!\tilde{k}_\phi}
\newcommand{\iitkp}{{}^{(2)}\!\tilde{k}_\phi}

\newcommand{\ajn}{{}^{(a)}\!j_0(\vec{r})}
\newcommand{\akn}{{}^{(a)}\!k_0\left(\vec{r}\right)}
\newcommand{\bkn}{{}^{(b)}\!k_0\left(\vec{r}\right)}
\newcommand{\bknn}{{}^{(b)}\!k_0\left(r\right)}
\newcommand{\btknn}{{}^{(b)}\!\tilde{k}_0\left(r\right)}
\newcommand{\aki}{{}^{(a)}\!k_1(\vec{r})}
\newcommand{\akp}{{}^{(a)}\!k_\phi(\vec{r})}
\newcommand{\akv}{\vec{k}_a(\vec{r})}

\newcommand{\aAn}{{}^{(a)}\!A_0(\vec{r})}
\newcommand{\pAn}{{}^{(\textrm{p})}\!A_0(\vec{r})}
\newcommand{\ppAn}{{}^{[\textrm{p}]}\!A_0(r)}
\newcommand{\iAn}{{}^{(1)}\!\!A_0(\vec{r})}
\newcommand{\iiAn}{ {}^{(2)}\!\! A_0 (\vec{r}) }

\newcommand{\aVHS}{{}^{(a)}V_{\textrm{HS}}(\vec{r})}
\newcommand{\iVHS}{{}^{(1)}V_{\textrm{HS}}(\vec{r})}
\newcommand{\iiVHS}{{}^{(2)}V_{\textrm{HS}}(\vec{r})}

\newcommand{\onp}{\omega_0^\textrm{(+)}}
\newcommand{\onm}{\omega_0^\textrm{(-)}}
\newcommand{\oip}{\omega_1^\textrm{(+)}}
\newcommand{\oim}{\omega_1^\textrm{(-)}}

\newcommand{\B}{\mathcal{B}}
\newcommand{\as}{\alpha_{\textrm{s}}\,}
\newcommand{\aB}{a_{\textrm{B}}\,}
\newcommand{\Bstar}{\overset{*}{B}}
\newcommand{\Gstar}{\overset{*}{G}}
\newcommand{\gstar}{\overset{*}{g}}
\newcommand{\CC}{\mathbb{C}}
\newcommand{\GG}{\textnormal I \! \Gamma}
\newcommand{\D}{\mathcal{D}}
\newcommand{\F}{\mathcal{F}}
\renewcommand{\H}{\mathcal{H}}
\newcommand{\I}{\mathcal{I}}
\newcommand{\J}{\mathcal{J}}
\newcommand{\LRST}{\mathcal{L}_{\textrm{RST}}}
\newcommand{\LD}{\mathcal{L}_{\textrm{D}}}
\newcommand{\LG}{\mathcal{L}_{\textrm{G}}}
\newcommand{\U}{\mathcal{U}}
\newcommand{\M}{\mathcal{M}}
\newcommand{\T}{\mathcal{T}}
\newcommand{\Mp}{M_\textrm{p}\,}
\newcommand{\Me}{M_\textrm{e}\,}
\newcommand{\MT}{M_\textrm{T}}
\newcommand{\tMT}{\tilde{M}_\textrm{T}}
\newcommand{\Mei}{M_\textrm{I}^\textrm{(e)}}
\newcommand{\Meii}{M_\textrm{II}^\textrm{(e)}}
\newcommand{\Mmi}{M_\textrm{I}^\textrm{(m)}}
\newcommand{\Mmii}{M_\textrm{II}^\textrm{(m)}}
\newcommand{\TT}{{}^{(\textrm{T})}T}
\newcommand{\DT}{{}^{(\textrm{D})}T}
\newcommand{\GT}{{}^{(\textrm{G})}T}
\newcommand{\heER}{\hat{E}_\textrm{R}^\textrm{(e)}}
\newcommand{\hmER}{\hat{E}_\textrm{R}^\textrm{(m)}}

\newcommand{\ED}{{E_{\textrm{D}}}}
\newcommand{\EG}{E_{\textrm{G}}}

\newcommand{\ET}{E_{\textrm{T}}}
\newcommand{\tET}{\tilde{E}_\textrm{T}}
\newcommand{\tETi}{\tilde{E}_\textrm{T}^{(|)}}
\newcommand{\tETii}{\tilde{E}_\textrm{T}^{(||)}}
\newcommand{\tEnT}{ {\tilde{E}^{(0)}_\textrm{T}} }
\newcommand{\tEnTmin}{ {\tilde{E}^{(0)}_\textrm{T,min}} }
\newcommand{\ETiii}{E_\textrm{T}^{(|||)}}
\newcommand{\Ekin}{{E_\textrm{kin}}}
\newcommand{\ES}{{E_{\textrm{S}}}}
\newcommand{\EW}{{E_\textrm{W}}}
\newcommand{\HS}{H_{\textrm{S}}}
\newcommand{\hHS}{\hat{H}_{\textrm{S}}}
\newcommand{\WRST}{W_{\textrm{RST}}}
\newcommand{\WS}{W_{\textrm{S}}}
\newcommand{\lG}{{\lambda_\textrm{G}}}
\newcommand{\lGe}{{\lambda_\textrm{G}}^{(\textrm{e})}}
\newcommand{\lGm}{{\lambda_\textrm{G}}^{(\textrm{m})}}
\newcommand{\lD}{{\lambda_\textrm{D}}}
\newcommand{\lS}{{\lambda_\textrm{S}}}
\newcommand{\ND}{{N_\textrm{D}}}
\newcommand{\NDn}{{N_\textrm{D}}^\textrm{(0)}}
\newcommand{\NGn}{{N_\textrm{G}}^\textrm{(0)}}
\newcommand{\NGe}{{N_\textrm{G}}^\textrm{(e)}}
\newcommand{\NGm}{{N_\textrm{G}}^\textrm{(m)}}
\newcommand{\LDk}{{\LD^{(\mathrm{kin})}}}
\newcommand{\LDe}{{\LD^{(\mathrm{e})}}}
\newcommand{\LDm}{{\LD^{(\mathrm{m})}}}
\newcommand{\LDM}{{\LD^{(\mathrm{M})}}}
\newcommand{\Tkin}{{T_\textrm{kin}}}
\newcommand{\oWD}{{\overset{\circ}{W}\!}_\textrm{D}}
\newcommand{\oWG}{{\overset{\circ}{W}\!}_\textrm{G}}
\newcommand{\oWRST}{{\overset{\circ}{W}\!}_\textrm{RST}}

\newcommand{\oWRSTe}{{\overset{\circ}{W}}{}^{\textrm{(e)}}_\textrm{RST}}

\newcommand{\Ph}{\Phi(\vec{r}_1,\vec{r}_2)}
\newcommand{\Z}{\mathcal{Z}}
%%% Local Variables: 
%%% mode: latex
%%% TeX-master: "main_minenergy"
%%% End: 

\title{\bf Principle of Minimal Energy\\ in\\ Relativistic Schr\"odinger Theory}
\author{M.\ Mattes and M.\ Sorg\\[1cm] II.\ Institut f\"ur Theoretische Physik der
Universit\"at Stuttgart\\ Pfaffenwaldring 57 \\ D 70550 Stuttgart, Germany\\Email:
sorg@theo2.physik.uni-stuttgart.de\\http://www.theo2.physik.uni-stuttgart.de/institut/sorg/publika.html\vspace{-2mm}}
\date{ }
\maketitle
\begin{abstract}
  The Hamilton-Lagrange action principle for Relativistic Schr\"o\-dinger Theory (RST) is
  converted to a variational principle (with constraints) for the stationary bound states.
  The groundstate energy is the minimally possible value of the corresponding energy
  functional and the relativistic energy eigenvalue equations do appear as the
  corresponding variational equations. The matter part of these eigenvalue equations is a
  relativistic generalization of the well-known Ritz principle in non-relativistic quantum
  mechanics which however disregards the dynamical character of the particle interactions.
  If the latter are included in the proposed principle of minimal energy for the bound
  states, one obtains a closed dynamical system for both matter and gauge fields. The new
  variational principle enables the development of variational techniques for solving
  approximately the energy eigenvalue equations. As a demonstration, the positronium
  groundstate is treated in great detail. Here a simple exponential trial function is
  sufficient in order to reproduce the (exact) result of conventional quantum mechanics
  where the relativistic and spin effects are neglected.

\end{abstract}

%%% Local Variables: 
%%% mode: latex
%%% TeX-master: "main_minenergy"
%%% End: 

\section{Introduction and Survey of Results}
\indent

The existence of an action principle is generally believed to be a very attractive feature
of any (quantum) field theory, especially concerning the modern gauge field theories. The
reason is that such an action principle does not only admit the deduction of the dynamical
equations of the theory but it provides also many additional advantages, e.g.\ deduction
of the conservation laws due to the symmetries of the theory (Noether theorem), canonical
and path integral quantization, coupling to other fields, etc. (see, e.g.,
ref.s~\cite{We,Gr}). In view of such a considerable relevance of the variational
principles, it does not appear as a surprise that there is an extended literature
concerning the mathematical structure of the variational principles themselves, e.g.\
ref.s \cite{La,Ru}.

One important aspect of these variational principles refers to the possibility of using
them for the construction of certain approximation techniques if exact solutions of the
dynamical equations cannot easily be found which will be mostly the case. The present
paper is also concerned with just this aspect of the variational principles, namely in the
context of the energy eigenvalue problems emerging within the framework of Relativistic
Schr\"odinger Theory (RST), a recently established theory of relativistic quantum
matter~\cite{BeSo,BMS,BeSo2}. More precisely, the intention of the paper aims at the
construction of a \emph{principle of minimal energy} for the bound RST field configurations
so that the groundstate carries the minimally possible value of the total field
energy~$\ET$; and furthermore the excited states represent the stationary
points~$(\delta\ET=0)$ of this energy functional~$\ET$. It should be rather evident
that the existence of such a minimal-energy principle is of invaluable practical
usefulness for the calculation of the (relativistic) energy levels of the bound systems
(i.e.\ atoms and molecules). The point here is that one is not necessarily forced to look
for the exact solutions of the eigenvalue equations but instead one can resort to the
invention of appropriate trial configurations of the RST fields with all those symmetries
(and other properties) being owned by the unknown exact solution, too. The principle of
minimal energy can then be used in order to find the trial configurations with minimal (or
stationary) energy which mostly is equivalent to a purely algebraic problem, namely the
determination of the ansatz parameters in the chosen trial functions.

For the sake of a simple demonstration and comparison with \emph{exactly} solvable examples of
the conventional theory, we select the positronium groundstate as a typical two-body
problem. Restricting ourselves here to the non-relativistic situation, one can easily show
that the choice of a simple trial function reproduces \emph{exactly} the positronium groundstate
energy as it is predicted by the conventional quantum mechanics.

These results are elaborated through the following sequence of intermediate steps:

\begin{center}
  \emph{\textbf{1.\ RST Eigenvalue Problem}}
\end{center}

As the point of departure for the construction of the desired variational principle, one
reconsiders the emergence of the typical energy eigenvalue problem within the framework of
RST. \textbf{Sect.~II} presents a brief sketch of the general theory for two oppositely
charged particles such as, e.g., hydrogen atom or positronium. The important point here is
that the subsequently defined energy functional~$\ET$ can be based upon the generally
accepted hypothesis of field energy which is concentrated in any relativistic field
configuration with energy-momentum density~$T_{\mu\nu}$, see equation (\ref{eq:II.53})
below. This object~$T_{\mu\nu}$ may be deduced from the corresponding RST
Lagrangean~$\LRST$ (\ref{eq:II.57}) in the usual way, i.e.\ via the standard Noether
theorem. Indeed, the existence of a Hamilton-Lagrange action principle
(\ref{eq:II.56a})-(\ref{eq:II.56b}) for RST is essential for the subsequent construction
of the desired minimal-energy principle for the bound states since this action principle
provides the possibility of introducing the \emph{Poisson identities} which then work as
constraints for the variational procedure (see the discussion of this point in the
preceding paper~\cite{BMS}).

Next, the stationary bound states are introduced in \textbf{Sect.~III} and lead
immediately to the \emph{mass eigenvalue equations} (\ref{eq:III.15}) and
(\ref{eq:III.17}) for the time-independent Dirac spinors~$\psi_a(\vec{r})$ of both
particles~$(a=1,2)$. The interesting point with these mass eigenvalue equations refers to
the fact that they can not only be deduced from the original Hamilton-Lagrange action
principle (\ref{eq:II.56a})-(\ref{eq:II.56b}) by means of the stationary ansatz
(\ref{eq:III.1a})-(\ref{eq:III.1b}) for both Dirac spinors, cf.\ (\ref{eq:III.14}) and
(\ref{eq:III.16}), but these mass eigenvalue equations can also be deduced from an
appropriately constructed mass functional~$\tMT c^2$ (\ref{eq:III.24}). It is true, this
mass functional can be understood to represent the immediate relativistic generalization
of the well-known Hartree-Ritz variational principle (\ref{eq:III.33a})-(\ref{eq:III.33b})
for factorized two-particle wave functions~$\Phi(\vec{r}_1,\vec{r}_2)$ (\ref{eq:III.36}),
but nevertheless~$\tMT c^2$ cannot be accepted to represent our wanted energy
functional~$\tET$ because this mass functional~$\tMT c^2$ suffers from the same
deficiencies as the conventional Ritz principle; namely, in the latter approach the
interaction between both particles is simply taken as the rigid Coulomb interaction (see
the conventional Hamiltonian~$\hHS$ (\ref{eq:III.34})) whereas it is well-known that the
gauge field, as the mediator of the relativistic interactions, must of course be treated
as a dynamical object obeying its own field equations (see the Maxwell equations
(\ref{eq:II.24})).

And furthermore, there is a second deficiency inherent in the Hartree-Ritz approach which
refers to the conventional dogma that wave functions should always be unique and
non-singular. However, the RST treatment of positronium in the preceding paper~\cite{BMS}
has shown that non-unique wave functions of the type (\ref{eq:III.60}) must be admitted.
Indeed, their use yields then a more realistic prediction of the positronium groundstate
energy as compared to the Hartree-Ritz-Schr\"odinger approach, see ref.~\cite{BMS}. The
conclusion from this is that in RST one must both admit a more general type of wave
function and treat the particle interaction as a proper dynamical object. Otherwise one
cannot expect to achieve a well-working principle of minimal energy which takes adequate
account of both the matter and gauge fields!

\begin{center}
  \emph{\textbf{2.\ Exotic Quantum States}}
\end{center}

The treatment of the positronium groundstate in the preceding paper~\cite{BMS}
demonstrates that the minimal value of the RST energy functional~$\ET$ cannot be reached
by admitting exclusively these non-singular and unique wave functions as they are usually
required by the Ritz-Hartree-Schr\"odinger approach in conventional quantum mechanics.
For instance, the requirement of physical equivalence of both positronium constituents
(i.e.\ positron and electron) entails that any of the two fermions has vanishing spin
component along the z-direction (equation (\ref{eq:III.61})) which is quite unusual for
fermionic particles. As a consequence of this unusual behavior of the fermions, several
other taboos of conventional quantum theory become broken, too:
\begin{itemize}
\item[\textbf{(i)}] The wave functions are singular at the origin~$(r=0)$ and along the
  whole z-axis from the very beginning, see equations (\ref{eq:III.62a})-(\ref{eq:III.62b})
  below; but nevertheless these singularities do not spoil the normalization conditions in
  the relativistic sense (\ref{eq:II.44}).
\item[\textbf{(ii)}] The wave functions become non-unique, e.g.\ in the sense of equation
  (\ref{eq:III.60}); but the observable physical densities (of charge, current,
  energy-momentum etc.) generated by these ambiguous wave functions are still unique and
  physically well-behaved.
\item[\textbf{(iii)}] The interaction potentials, generated by the exotic states, are
  singular at the origin but less singular as the standard Coulomb potential so that their
  field energy is kept finite and can thus enter the wanted energy functional~$\ET$
  without causing infinities, see equations (\ref{eq:III.76})-(\ref{eq:III.77}) below.
\item[\textbf{(iv)}] The magnetic moment carried by the bound matter fields amounts only
  to half a Bohr magneton~$\mu_{\mathrm{B}}$, see the asymptotic form of the magnetic
  potential in equation (\ref{eq:IV.48}) below.
\end{itemize}

\begin{center}
  \emph{\textbf{3.\ Positronium Groundstate}}
\end{center}

As a concrete demonstration, all this theoretical structure is evoked in order to
calculate the positronium groundstate energy~$E_0$. First, in the absence of an exact
solution to the corresponding RST eigenvalue problem (consisting of the coupled set of
relativistic mass eigenvalue equations (\ref{eq:III.66a})-(\ref{eq:III.66d}) and Poisson
equations (\ref{eq:III.47a})-(\ref{eq:III.47d})) one resorts to a self-suggesting
variational technique based upon the constructed energy functional~$\tET$ (\ref{eq:IV.10})
which, however, is applied in this paper only in its non-relativistic
approximation~$\tEnT$ (\ref{eq:IV.26}). This means that one has to guess a trial function
as realistic as possible (see the simple exponential wave amplitude~$\tilde{R}(r)$
(\ref{eq:V.5})) and substitutes this into the energy functional~$\tEnT$ (\ref{eq:IV.26}).
This energy functional is additively composed of the kinetic energies~$\Ekin(a)$ of both
particles (\ref{eq:IV.20a})-(\ref{eq:IV.20b}) plus their electrostatic interaction
energy~$\heER$ (\ref{eq:IV.22a}); the magnetic interaction energy~$\hmER$
(\ref{eq:IV.22b}) is first neglected and treated afterwards as a small perturbation of the
electric effects.

The energy functional~$\tEnT$ (\ref{eq:IV.26}) contains also two constraints which have to
be respected for the deduction of the mass eigenvalue and Poisson equations as the
variational equations due to that functional (i.e.~$\delta\tEnT=0$). The first constraint
refers to the wave function normalization (as shown, e.g., by equation (\ref{eq:IV.47}))
and is automatically satisfied by our ansatz (\ref{eq:V.9a}). However, the second
constraint refers to the electric Poisson identities, such as (\ref{eq:III.58}), and
requires a more subtle argument: if one wishes to have the Poisson constraints also
automatically satisfied by the trial functions one first has to solve the corresponding
Poisson equations; i.e.\ equation (\ref{eq:III.72}) for the present situation. But if all
constraints are thus satisfied automatically by our trial function~$\tilde{R}(r)$, one
substitutes this into the energy functional~$\tEnT$ (\ref{eq:IV.26}) and obtains an
ordinary function~$\tEnT(r_*)$ of the ansatz parameter~$r_*$, i.e.\ equation
(\ref{eq:V.7}) which according to the \emph{principle of minimal energy} adopts the
groundstate energy~$E_0$ (\ref{eq:V.8}) as its minimal value.  This just coincides with
the corresponding prediction (\ref{eq:V.1}) of conventional quantum mechanics.  However,
observe here that this groundstate energy~$E_0$ (\ref{eq:V.1}) owns the status of
\emph{exactness} within the framework of the conventional theory, whereas in RST it
appears as an \emph{approximation} (even if all the relativistic effects including
magnetism are disregarded) since our trial function~$\tilde{R}(r)$ (\ref{eq:V.5}) is
surely not the exact solution of the non-relativistic RST eigenvalue problem in the
electrostatic approximation.

Finally, the magnetic interaction energy~$\hmER$ (\ref{eq:IV.56}) is estimated in the
lowest-order of approximation, equation (\ref{eq:V.16}) below. It turns out that the RST
prediction for the hyperfine splitting of the positronium groundstates~${}^1 S_0$
and~${}^3 S_1$ amounts to only~$1,47\cdot 10^{-4} [eV]$, whereas the experimental value
is~$8,41\cdot 10^{-4} [eV]$~\cite{LeWe}. Thus this lowest-order RST prediction shows that
for the hyperfine splitting one needs a better trial function.

%%% Local Variables: 
%%% mode: latex
%%% TeX-master: "main_minenergy"
%%% End: 

\section{Two-Fermion Systems in RST}
\indent

In order to introduce the relevant notation, a brief sketch of the general two-particle
theory is presented first so that the characteristic dynamical structure becomes obvious:
matter dynamics, Hamiltonian dynamics, gauge field dynamics, \emph{action principle}, and the
associated conservation laws (for a more detailed presentation of the RST dynamics, see
the preceding papers \cite{BeSo}-\cite{BeSo2}). It is true, the existence of an action
principle is common to almost all of the successful field theories, but a pleasant feature
of the present RST dynamics refers to the fact that its action principle can be converted
to a \emph{principle of minimal energy} for the bound systems. This will subsequently be
exploited in order to compute approximately the positronium groundstate energy.

\begin{center}
  \emph{\textbf{A.\ Matter Dynamics}}
\end{center}

The central equation of motion for matter is the \textbf{R}elativistic
\textbf{S}chr\"odinger \textbf{E}quation~(RSE)
\begin{equation}
  \label{eq:II.1}
  i\hbar c\D_\mu\Psi = \H_\mu\Psi\ ,
\end{equation}
or if matter is to be described by an intensity matrix~$\I$ in place of a pure
state~$\Psi$, one applies the \textbf{R}elativistic von \textbf{N}eumann
\textbf{E}quation~(RNE)
\begin{equation}
  \label{eq:II.2}
  \D_\mu\I=\frac{i}{\hbar c}\left(\I\bar{\H}_\mu - \H_\mu\I \right)\ .
\end{equation}
In the present paper, we will exclusively deal with pure two-particle states~$\Psi$ which
in RST are always the direct (Whitney) sum of the one-particle states $\psi_a\ (a=1,2)$, i.e.
\begin{equation}
  \label{eq:II.3}
  \Psi(x) = \psi_1(x)\oplus\psi_2(x)\ .
\end{equation}
Here the one-particle states~$\psi_a(x)$ are four-component Dirac spinor fields so that
the two-particle wave function~$\Psi(x)$ may be understood as a section of a complex
vector bundle over space-time as the base space with typical fibre~$\CC^8$.

Both particles are interacting with each other via the principle of minimal coupling, i.e.\ the
gauge-covariant derivative in the RSE~(\ref{eq:II.1}) is defined by means of the gauge
potential~$\A_\mu$ (\emph{bundle connection}) in the usual way as
\begin{equation}
  \label{eq:II.4}
  \D_\mu\Psi = \partial_\mu\Psi+\A_\mu\Psi\ ,
\end{equation}
or, resp., in component form
\begin{equation}
  \label{eq:II.5}
  \D_\mu\Psi = \left(D_\mu\psi_1\right)\oplus\left(D_\mu\psi_2\right)\ .
\end{equation}
Here the gauge-covariant derivatives of the one-particle states~$\psi_a(x)$ are given by
\begin{subequations}
  \begin{align}
    \label{eq:II.6a}
  D_\mu\psi_1 &= \partial_\mu\psi_1 - i A^2_\mu\psi_1 - iB_\mu\psi_2\\
    \label{eq:II.6b}
  D_\mu\psi_2 &= \partial_\mu\psi_2 - i A^1_\mu\psi_2 - i\Bstar_\mu\psi_1\ ,  
  \end{align}
\end{subequations}
provided the bundle connection~$\A_\mu$ takes its values in the four-dimensional Lie
algebra~$\U(2)$ of the unitary group~U(2) (\emph{structure group}) and is decomposed with
respect to a suitable basis of generators~$\{\tau_a,\chi,\bar{\chi}\}$ as follows:
\begin{equation}
  \label{eq:II.7}
  \A_\mu = \sum_{a=1}^2 {A^a}_\mu\tau_a + B_\mu\chi - \Bstar_\mu\bar{\chi}\ .
\end{equation}
The (real-valued) \emph{electromagnetic potentials}~${A^a}_\mu$ do mediate the
electromagnetic interactions between both particles; and similarly the (complex-valued)
\emph{exchange potentials}~$B_\mu$ do mediate the exchange interactions which thus are treated in
RST as real forces on the same footing as their electromagnetic counterparts. However the
exchange forces (due to~$B_\mu$) can be active exclusively among \emph{identical}
particles and must vanish~$(B_\mu\equiv 0)$ for \emph{non-identical} particles (see
refs.s~\cite{BeSo,BeSo2}). Since we restrict ourselves in the present paper to a system of two
\emph{oppositely} charged particles with different or identical masses~$\Mp$ and~$\Me$, resp.,
the exchange forces must therefore be zero and consequently the covariant
derivatives~(\ref{eq:II.6a})-~(\ref{eq:II.6b}) simplify to
\begin{subequations}
  \begin{align}
    \label{eq:II.8a}
    D_\mu\psi_1 &= \partial_\mu\psi_1 - i A^2_\mu\psi_1\\
    \label{eq:II.8b}
    D_\mu\psi_2 &= \partial_\mu\psi_2 - i A^1_\mu\psi_2\ .
  \end{align}
\end{subequations}

\pagebreak
\begin{center}
  \emph{\textbf{B.\ Hamiltonian Dynamics}}
\end{center}

The \emph{Hamiltonian}~$\H_\mu$, occurring in the RSE~(\ref{eq:II.1}) or in the
RNE~(\ref{eq:II.2}), takes its values in the general linear algebra~$\mathcal{GL}(2,\CC)$
and is itself a dynamical object which is to be determined from its field equations, i.e.\
the \emph{integrability condition}
\begin{equation}
  \label{eq:II.9}
  \D_\mu\H_\nu -\D_\nu\H_\mu + \frac{i}{\hbar c}\left[\H_\mu,\H_\nu \right] = i\hbar c
  \F_{\mu\nu}  
\end{equation}
and the \emph{conservation equation}
\begin{equation}
  \label{eq:II.10}
  \D^\mu\H_\mu - \frac{i}{\hbar c}\H^\mu\H_\mu = -i\hbar c \left[\left(\frac{\M c}{\hbar}\right)^2 + 
    \Sigma^{\mu\nu}\F_{\mu\nu}\right]\ .
\end{equation}
The integrability condition~(\ref{eq:II.9}) contains the curvature~$\F_{\mu\nu}$ of the
bundle connection~$\A_\mu$~(\ref{eq:II.7})
\begin{equation}
  \label{eq:II.11}
  \F_{\mu\nu} \doteqdot \nabla_\mu\A_\nu -\nabla_\nu\A_\mu + \left[\A_\mu,\A_\nu\right]
\end{equation}
and guarantees the validity of the bundle identities
\begin{subequations}
  \begin{align}
    \label{eq:II.12a}
    \left[\D_\mu\D_\nu - \D_\nu\D_\mu\right]\Psi &= \F_{\mu\nu}\Psi\\
    \label{eq:II.12b}
    \left[\D_\mu\D_\nu - \D_\nu\D_\mu\right]\I &= \left[\F_{\mu\nu},\I\right]\ .
  \end{align}
\end{subequations}

The conservation equation~(\ref{eq:II.10}) contains the \emph{mass operator}~$\M$ and the
Spin(1,3) generators~$\Sigma_{\mu\nu}$
\begin{equation}
  \label{eq:II.13}
  \Sigma_{\mu\nu} = \frac{1}{4}\left[\GG_\mu,\GG_\nu\right]
\end{equation}
which both are assumed to be covariantly constant
\begin{subequations}
  \begin{align}
    \label{eq:II.14a}
    \D_\mu\M &\equiv 0\\
    \label{eq:II.14b}
    \D_\mu\Sigma_{\lambda\nu} &\equiv 0\ .
  \end{align}
\end{subequations}
The latter constancy condition~(\ref{eq:II.14b}) may be traced back to the covariant
constancy of the \emph{total velocity operator}~$\GG_\mu$
\begin{equation}
  \label{eq:II.15}
  \D_\lambda\GG_\mu\equiv 0\ ,
\end{equation}
where~$\GG_\mu$ are the direct sum of the one-particle Dirac matrices~$\gamma_\mu$
\begin{equation}
  \label{eq:II.16}
  \GG_\mu=(-\gamma_\mu)\oplus\gamma_\mu
\end{equation}
and therefore can be taken as the generators of the required eight-dimensional
representation of the Clifford algebra~$\CC(1,3)$, i.e.
\begin{equation}
  \label{eq:II.17}
  \GG_\mu\GG_\nu + \GG_\nu\GG_\mu = 2g_{\mu\nu}\cdot\mathbf{1}_{(8)}\ .
\end{equation}
Observe here that the arrangement of the plus and minus signs in the direct
sum~(\ref{eq:II.16}) displays the opposition of both particle charges (positive charge of
the first particle and negative charge of the second particle, by convention). 

The conservation equation~(\ref{eq:II.10}) is needed for the deduction of the conservation
laws from the RST dynamics (see below) and admits an equivalent algebraic formulation:
\begin{equation}
  \label{eq:II.18}
  \GG^\mu\H_\mu = \bar{\H}_\mu\GG^\mu = \M c^2\ .
\end{equation}
This can be used in order to eliminate the Hamiltonian~$\H_\mu$ by recasting the
RSE~(\ref{eq:II.1}) into the two-particle \textbf{D}irac \textbf{E}quation~(DE)
\begin{equation}
  \label{eq:II.19}
  i\hbar\GG^\mu\D_\mu\Psi = \M c\Psi\ .
\end{equation}
In component form, this equation reads
\begin{subequations}
  \begin{align}
    \label{eq:II.20a}
    i\hbar\gamma^\mu D_\mu\psi_1 &= -\Mp c \psi_1\\*
    \label{eq:II.20b}
    i\hbar\gamma^\mu D_\mu\psi_2 &= \Me c \psi_2\ ,
  \end{align}
\end{subequations}
provided the (covariantly constant) mass operator~$\M$ is written as
\begin{equation}
  \label{eq:II.21}
  \M = i\sum_{a=1}^{2}M^a\tau_a
\end{equation}
with
\begin{subequations}
  \begin{align}
  \label{eq:II.22a}
  M^1 &\doteqdot \Me\\*
  \label{eq:II.22b}
  M^2 &\doteqdot \Mp\ ,
  \end{align}
\end{subequations}
where~$\Mp$ and~$\Me$ are denoting the rest mass of the positively and negatively
charged particle, resp. For the case of pure states, one can eliminate the
Hamiltonian~$\H_\mu$ also by differentiating once more the RSE~(\ref{eq:II.1}) and
substituting therein the derivative of~$\H_\mu$ from the original conservation
equation~(\ref{eq:II.10}) which yields a second-order equation of the
\textbf{K}lein-\textbf{G}ordon type~(KGE):
\begin{equation}
  \label{eq:II.23}
  \D^\mu\D_\mu\Psi + \left(\frac{\M c}{\hbar}\right)^2\Psi =
  -\Sigma^{\mu\nu}\F_{\mu\nu}\Psi\ .
\end{equation}
However, subsequently we will prefer to deal with the first order equation~(\ref{eq:II.19}).
\pagebreak

\begin{center}
  \emph{\textbf{C.\ Gauge Field Dynamics}}
\end{center}

In order to close the RST dynamics, one finally has to specify some field equation for the
bundle connection~$\A_\mu$. Our choice is the non-Abelian \emph{Maxwell equation}
\begin{equation}
  \label{eq:II.24}
  \D^\mu\F_{\mu\nu} = -4\pi i \as \J_\nu
\end{equation}
where the \emph{current operator}~$\J_\mu$ may be thought to decompose with respect to
the structure algebra basis~$\{\tau_\alpha,\ \alpha=1\ldots 4 \}=\{\tau_a,\chi,\bar{\chi} \}$
as follows:
\begin{equation}
  \label{eq:II.25}
  \J_\mu=i{j^\alpha}_\mu\tau_\alpha=i\left({j^1}_\mu\tau_1+{j^2}_\mu\tau_2+g_\mu\chi-\gstar_\mu\bar{\chi} \right)\ . 
\end{equation}
Here, the \emph{Maxwell currents}~${j^a}_\mu (a=1,2)$ generate the electromagnetic
potentials~${A^a}_\mu$~(\ref{eq:II.7}) which is seen by explicitly writing down the
electromagnetic part of the general Maxwell equations~(\ref{eq:II.24}) in component
form~$(a=1,2)$
\begin{equation}
  \label{eq:II.26}
  D^\mu {F^a}_{\mu\nu} = 4\pi\as{j^a}_\nu\ .
\end{equation}
The \emph{exchange currents}~${j^3}_\mu\doteqdot g_\mu$ and~${j^4}_\mu\doteqdot
-\gstar_\mu$ do generate the exchange potentials~$B_\mu$ and~$\Bstar_\mu$~(\ref{eq:II.7});
but since we are dealing here exclusively with non-identical particles the exchange
potentials~$B_\mu,\Bstar_\mu$ must be put to zero so that the Maxwell
equations~(\ref{eq:II.26}) become \emph{Abelian}:
\begin{subequations}
  \begin{align}
    \label{eq:II.27a}
    \nabla^\mu {F^1}_{\mu\nu} &= 4\pi\as{j^1}_\nu\\*
    \label{eq:II.27b}
    \nabla^\mu {F^2}_{\mu\nu} &= 4\pi\as{j^2}_\nu\ .
  \end{align}
\end{subequations}
The formal reason for this is that the bundle curvature~$\A_\mu$~(\ref{eq:II.7}) and its
curvature~$\F_{\mu\nu}$~(\ref{eq:II.11})
\begin{equation}
  \label{eq:II.28}
  \F_{\mu\nu}=\sum_{a=1}^2{F^a}_{\mu\nu} \tau_a+G_{\mu\nu}\chi-\Gstar_{\mu\nu}\bar{\chi}
\end{equation}
become projected onto the Abelian subalgebra~$\U(1)\oplus\U(1)$ when the exchange
fields~$B_\mu,G_{\mu\nu}$ are put to zero.

\begin{center}
  \emph{\textbf{D.\ Conservation Laws}}
\end{center}

The right choice of the gauge field dynamics is not a trivial thing because it must be
compatible with the already fixed matter dynamics (for both the pure states and the
mixtures). However, this desired compatibility of our choice can be verified in the general
case by the following arguments: First, the generally valid bundle identity
\begin{equation}
  \label{eq:II.29}
  \D^\mu\D^\nu \F_{\mu\nu}\equiv 0\ ,
\end{equation}
when applied to the Maxwell equations~(\ref{eq:II.24}), yields the following source
equation for the current operator~$\J_\mu$
\begin{equation}
  \label{eq:II.30}
  \D^\mu\J_\mu\equiv 0\ ,
\end{equation}
or in component form
\begin{equation}
  \label{eq:II.31}
  D^\mu {j^\alpha}_\mu\equiv 0\ .
\end{equation}
This means that the two-particle Maxwell currents~${j^\alpha}_\mu (\alpha=1,\ldots 4)$
must be constructed in terms of the two-particle wave function~$\Psi$ in such a way that
the covariant source equations~(\ref{eq:II.31}) do actually hold just as a consequence of the
RST dynamics!

This compatibility requirement can be satisfied by first constructing the \emph{RST
  currents}~$j_{\alpha\mu}$ through
\begin{equation}
  \label{eq:II.32}
  j_{\alpha\mu} \doteqdot \bar{\Psi} v_{\alpha\mu}\Psi
\end{equation}
with the \emph{velocity operators}~$v_{\alpha\mu}$ being defined through the following
anticommutators
\begin{equation}
  \label{eq:II.33}
  v_{\alpha\mu}=\frac{i}{2}\left\{\tau_\alpha,\GG_\mu\right\}\ .
\end{equation}
Indeed, one can easily show that these RST currents~$j_{\alpha\mu}$~(\ref{eq:II.32}) do
obey the source equations
\begin{equation}
  \label{eq:II.34}
  D^\mu j_{\alpha\mu}\equiv 0\ ,
\end{equation}
provided the wave function~$\Psi$ (or intensity matrix~$\I$, resp.) does satisfy the
RSE~(\ref{eq:II.1}) (or the RNE~(\ref{eq:II.2}), resp.). However, observe here that the
RST currents~$j_{\alpha\mu}$~(\ref{eq:II.32}) cannot a priori identified with the Maxwell
currents~${j^\alpha}_\mu$~(\ref{eq:II.25}) generating the gauge
potentials~${A^\alpha}_\mu$ via the Maxwell equations~(\ref{eq:II.26}). Consequently, there
must be established some link between the Maxwell currents~${j^\alpha}_\mu$ and RST
currents~$j_{\alpha\mu}$ in such a way that both source equations~(\ref{eq:II.31})
and~(\ref{eq:II.34}) are simultaneously valid! This requirement can be satisfied by
conceiving~${j^\alpha}_\mu$ and~$j_{\alpha\mu}$ as contra- and covariant versions of
one and the same object; namely by introducing a covariantly constant fibre
metric~$K_{\alpha\beta}$ for the associated Lie algebra bundle
\begin{equation}
  \label{eq:II.35}
  D_\lambda K_{\alpha\beta}\equiv 0\ ,
\end{equation}
and then putting
\begin{subequations}
  \begin{align}
    \label{eq:II.36a}
    {j^\alpha}_\mu &= K^{\alpha\beta}j_{\beta\mu}\\*
    \label{eq:II.36b}
    j_{\alpha\mu} &= K_{\alpha\beta}{j^\beta}_\mu\ .
  \end{align}
\end{subequations}

Actually, such a \emph{compatibility tensor}~$K_{\alpha\beta}$ can be found:
\begin{equation}
  \label{eq:II.37}
  K_{\alpha\beta} =
  C_1\tr\tau_\alpha\cdot\tr\tau_\beta+C_2\tr\left(\tau_\alpha\cdot\tau_\beta\right)\ ,
\end{equation}
where~$C_1$ and~$C_2$ are constants which have to be chosen in such a way that the
following constraint for the currents holds:
\begin{equation}
  \label{eq:II.38}
  \sum_{a=1}^2 {j^a}_\mu = -\sum_{a=1}^2 j_{a\mu} \doteqdot -j_\mu\ .
\end{equation}
Here the \emph{total current}~$j_\mu$ of the two-particle system appears as the sum of the
Maxwell (or RST) currents and acts as the source of the \emph{total electromagnetic
field}~$F_{\mu\nu}$ 
\begin{equation}
  \label{eq:II.39}
  F_{\mu\nu} \doteqdot {F^1}_{\mu\nu} + {F^2}_{\mu\nu}\ ,
\end{equation}
i.e.\ one easily deduces from the Abelian Maxwell
equations~(\ref{eq:II.27a})-(\ref{eq:II.27b}) the \emph{total Maxwell equation}
\begin{equation}
  \label{eq:II.40}
  \nabla^\mu F_{\mu\nu} = -4\pi\as j_\nu\ .
\end{equation}

Moreover, an immediate consequence of this Maxwell equation is the \emph{continuity
  equation} for the total current~$j_\mu$
\begin{equation}
  \label{eq:II.41}
  \nabla^\mu j_\mu \equiv 0\ .
\end{equation}
Therefore the total charge~$z$ may be defined through
\begin{equation}
  \label{eq:II.42}
  z=\int_{(S)} j_\mu dS^\mu\ ,
\end{equation}
which is independent of the chosen hypersurface~$(S)$ but must of course be zero because
we are dealing with opposite charges. This may be realized more clearly by expressing the
RST currents~$j_{a\mu}$ through the \emph{Dirac currents}~$k_{a\mu}$
\begin{subequations}
  \begin{align}
    \label{eq:II.43a}
    j_{1\mu}=k_{2\mu} \doteqdot \bar{\psi}_2\gamma_\mu\psi_2\\*
    \label{eq:II.43b}
    j_{2\mu}=-k_{1\mu} \doteqdot - \bar{\psi}_1\gamma_\mu\psi_1\ .
  \end{align}
\end{subequations}
Thus, since anyone of the two particles is assumed to carry just one charge unit, one will
apply the following normalization of the wave functions for the stationary bound
states~$(a=1,2)$ 
\begin{equation}
  \label{eq:II.44}
  \int d^3\vec{r}\,\,\akn = 1\ ,\\*
\end{equation}
where the hypersurface~$(S)$ in~(\ref{eq:II.42}) is taken as a time
slice~$(t=\mathrm{const.})$ of space-time; and the stationary form of the Dirac
currents~$k_{a\mu}$~(\ref{eq:II.43a})-(\ref{eq:II.43b}) is of course
\begin{equation}
  \label{eq:II.45}
  k_{a\mu}(x) = \left(\akn;-\akv \right)\ .
\end{equation}
Obviously the total charge~$z$~(\ref{eq:II.42}) becomes actually zero, namely by simply
observing the sum requirement~(\ref{eq:II.38}) and applying the normalization
conditions~(\ref{eq:II.44}) together with the relationship
(\ref{eq:II.43a})-(\ref{eq:II.43b}) between the RST and Dirac currents.

It is very instructive to consider the local charge conservation~(\ref{eq:II.41}) also
from an other viewpoint: The total current~$j_\mu$ may be defined alternatively through
\begin{equation}
  \label{eq:II.46}
  j_\mu = \bar{\Psi}\GG_\mu\Psi\ .
\end{equation}
Carrying here out the differentiation process~(\ref{eq:II.41}) and using the Dirac
equation~(\ref{eq:II.19}) together with the covariant constancy of the total velocity
operator~$\GG_\mu$~(\ref{eq:II.15}) actually yields just the total charge conservation
law~(\ref{eq:II.41}). A similar procedure does apply also to the local energy-momentum
conservation
\begin{equation}
  \label{eq:II.47}
  \nabla^\mu\,\TT_{\mu\nu}\equiv 0\ ,
\end{equation}
where~$\TT_{\mu\nu}$ is the \emph{total energy-momentum density} of the field
configuration and is composed of a matter part~$\DT_{\mu\nu}$ and a gauge field
part~$\GT_{\mu\nu}$
\begin{equation}
  \label{eq:II.48}
  \TT_{\mu\nu} = \DT_{\mu\nu}+\GT_{\mu\nu}\ .
\end{equation}
The interesting point here is that the individual sources of the partial
densities~$\DT_{\mu\nu}$ and~$\GT_{\mu\nu}$ turn out as the Lorentz forces which are
mutually annihilating:
\begin{equation}
  \label{eq:II.49}
  \nabla^\mu\,\DT_{\mu\nu} = - \nabla^\mu\,\GT_{\mu\nu} = \hbar c{F^\alpha}_{\mu\nu}{j_\alpha}^\mu
\end{equation}
so that the local law~(\ref{eq:II.47}) can be true. However the crucial condition for this
pleasant result is, that the RST dynamics (i.e.\ matter and gauge field dynamics) is chosen
as described above and that the partial densities are defined as follows:
\begin{subequations}
  \begin{align}
    \label{eq:II.50a}
    \GT_{\mu\nu} &= \frac{\hbar c}{4\pi\as}K_{\alpha\beta}\left({F^\alpha}_{\mu\lambda}
      {{F^\beta}_\nu}^\lambda -
      \frac{1}{4}g_{\mu\nu}{F^\alpha}_{\sigma\lambda}F^{\beta\sigma\lambda}\right)\\*
    \label{eq:II.50b}
    \DT_{\mu\nu} &= \bar{\Psi}\T_{\mu\nu}\Psi\ ,
  \end{align}
\end{subequations}
with the \emph{energy-momentum operator}~$\T_{\mu\nu}$ being given in terms of the
Hamiltonian~$\H_\mu$ and total velocity operator~$\GG_\mu$ as
\begin{equation}
  \label{eq:II.51}
  \T_{\mu\nu}=\frac{1}{4}\left(\GG_\mu\H_\nu+\bar{\H}_\nu\GG_\mu+\GG_\nu\H_\mu+\bar{\H}_\mu\GG_\nu \right)\ .
\end{equation}

If matter can be described by a pure state~$\Psi$ (in place of a mixture) so that the
matter density is given in terms of~$\Psi$ by~(\ref{eq:II.50b}), then the
Hamiltonian~$\H_\mu$ can again be eliminated from the matter density~$\DT_{\mu\nu}$ by
means of the DE~(\ref{eq:II.19}) which yields
\begin{equation}
  \label{eq:II.52}
  \DT_{\mu\nu} = \frac{i\hbar c}{4}\left[\bar{\Psi}\GG_\mu \left(\D_\nu\Psi\right) -
    \left(\D_\nu\bar{\Psi}\right) \GG_\mu\Psi + \bar{\Psi}\GG_\nu \left(\D_\mu\Psi\right) -  
    \left(\D_\mu\bar{\Psi}\right)\GG_\nu\Psi \right]\ .
\end{equation}
Clearly, the energy-momentum density~$\TT_{\mu\nu}$ is the crucial object for testing the
practical usefulness of the theory, because the corresponding energy content~$\ET$ of the
field configuration is given by the spatial integral of the time
component~$\TT_{00}(\vec{r})$, i.e.
\begin{equation}
  \label{eq:II.53}
  \ET = \int d^3\vec{r}\,\TT_{00}(\vec{r})\ .
\end{equation}
Since the density~$\TT_{\mu\nu}$~(\ref{eq:II.48}) appears as the sum of a matter and gauge field part, the
same must hold also for the total energy~$\ET$~(\ref{eq:II.53})
\begin{equation}
  \label{eq:II.54}
  \ET = \ED + \EG\ ,
\end{equation}
with the individual contributions being defined in a self-evident way as
\begin{subequations}
  \begin{align}
    \label{eq:II.55a}
    \ED &= \int d^3\vec{r}\, \DT_{00}(\vec{r})\\*
    \label{eq:II.55b}
    \EG &= \int d^3\vec{r}\, \GT_{00}(\vec{r})\ .
  \end{align}
\end{subequations}

Subsequently we will clarify the question whether for the groundstate of the stationary
two-particle systems the energy functional~(\ref{eq:II.53}) adopts its minimally possible
value~($\leadsto$ \emph{principle of minimal energy}).

\begin{center}
  \emph{\textbf{E.\ Action Principle}}
\end{center}

The conservation laws for charge~(\ref{eq:II.41}) and energy-momentum~(\ref{eq:II.47}) can
be directly deduced from the general RST dynamics, but a more elegant method is provided
by the Noether theorem~\cite{SSMS}. For the latter method one needs an action principle
\begin{subequations}
  \begin{align}
    \label{eq:II.56a}
    \delta\WRST &= 0\\*
    \label{eq:II.56b}
    \WRST &= \int d^4x\,\LRST[\Psi,\A_\mu]
  \end{align}
\end{subequations}
from which both the matter dynamics~(\ref{eq:II.19}) and the gauge field
dynamics~(\ref{eq:II.24}) may be deduced by the usual variational methods. The
corresponding RST Lagrangean~$\LRST$ has been specified as a sum of the matter part~$\LD$
and gauge field part~$\LG$~\cite{SSMS}
\begin{equation}
  \label{eq:II.57}
  \LRST[\Psi,\A_\mu] = \LD[\Psi] + \LG[\A_\mu]\ ,
\end{equation}
where the matter part is given by
\begin{equation}
  \label{eq:II.58}
  \LD[\Psi]=\frac{i\hbar c}{2}\left[\bar{\Psi}\GG^\mu\left(\D_\mu\Psi\right)-\left(\D_\mu\bar{\Psi}\right) \GG^\mu\Psi
    \right] - \bar{\Psi}\M c^2 \Psi
\end{equation}
and the gauge field part by
\begin{equation}
  \label{eq:II.59}
  \LG[\A_\mu] = \frac{\hbar c}{16\pi\as}K_{\alpha\beta}{F^\alpha}_{\mu\nu}F^{\beta\mu\nu}\ .
\end{equation}
Concerning the latter part~(\ref{eq:II.59}), observe here that the bundle
curvature $\F_{\mu\nu}$~(\ref{eq:II.11}) takes its values in the
subalgebra~$\U(1)\oplus\U(1)$ because we are dealing with non-identical particles; and
thus the gauge field Lagrangean becomes reduced to
\begin{equation}
  \label{eq:II.60}
  \LG[\A_\mu]=\frac{\hbar c}{16\pi\as}\sum_{a,b=1}^2 K_{ab}{F^a}_{\mu\nu}F^{b\mu\nu}\ .
\end{equation}

If the self-interactions are neglected, the fibre submetric~$K_{ab}$ is of a very simple
shape~\cite{BeSo}
\begin{equation}
  \label{eq:II.61}
  \left\{K_{ab}\right\} = 
  \begin{pmatrix}
    0  & -1\\*
    -1 & 0
  \end{pmatrix}\ .
\end{equation}
Thus, from the formal point of view,~$\LG[\A_\mu]$ describes the interaction of the two
gauge field modes~${F^a}_{\mu\nu}\ (a,b=1,2)$, i.e.
\begin{equation}
  \label{eq:II.62}
  \LG[\A_\mu]=\frac{\hbar c}{4\pi\as}\left( \vec{E}_1 \sdot \vec{E}_2 - \vec{H}_1
    \sdot \vec{H}_2 \right)\ ,
\end{equation}
provided the four-tensors~${F^a}_{\mu\nu}$ are splitted into their space and time
components~$\vec{E}_a,\vec{H}_a$ as usual
\begin{subequations}
  \begin{align}
    \label{eq:II.63a}
    \vec{E}_a &= \left\{{}^{(a)}E^j \right\} \doteqdot \left\{{F^a}_{0j} \right\}\\*
   \label{eq:II.63b} 
    \vec{H}_a &= \left\{{}^{(a)}H^j \right\} \doteqdot \left\{ \frac{1}{2}{\varepsilon^{jk}}_l
      F^a{}_k{}^l \right\}\ .
  \end{align}
\end{subequations}

But once the Lagrangean has been specified, it is an easy exercise to deduce both the RST
matter dynamics~(\ref{eq:II.19}) and the gauge field dynamics~(\ref{eq:II.24}) from the
action principle~(\ref{eq:II.56a})-(\ref{eq:II.56b}) as the corresponding Euler-Lagrange
equations. Furthermore, the considered conservation laws of charge~(\ref{eq:II.30}) and
energy-momentum~(\ref{eq:II.47}) are just those which are predicted by the Noether
formalism, see ref.~\cite{SSMS}. 
%%% Local Variables: 
%%% mode: latex
%%% TeX-master: "minenergie"
%%% End: 

\section{Stationary Bound Systems}
\indent

In order to make the proposed exercise with the Euler-Lagrange equations somewhat more
instructive and fruitful, one may immediately pass over to the stationary systems which are
defined through the usual product ansatz for the wave functions
\begin{subequations}
  \begin{align}
    \label{eq:III.1a}
    \psi_1(\vec{r},t) &= \exp\left(-i\frac{M_1 c^2}{\hbar}t \right)\cdot\psi_1(\vec{r})\\*
    \label{eq:III.1b}
    \psi_2(\vec{r},t) &= \exp\left(-i\frac{M_2 c^2}{\hbar}t \right)\cdot\psi_2(\vec{r})\ ,
  \end{align}
\end{subequations}
whereas the electromagnetic potentials become time-independent:
\begin{gather}
  \label{eq:III.2}
  {A^a}_\mu = \left\{ \aAn;-\vec{A}_a(r) \right\}\\*
  \left(a=1,2\right)\ .\notag
\end{gather}
This time-independence obviously does then apply also for the Dirac
currents~$k_{a\mu}$~(\ref{eq:II.43a})-(\ref{eq:II.43b}), see equation~(\ref{eq:II.45}).

\begin{center}
  \emph{\textbf{A.\ Mass Eigenvalue Equations}}
\end{center}

The \emph{mass eigenvalues}~$M_a\ (a=1,2)$, occurring in the stationary
ansatz (\ref{eq:III.1a})-(\ref{eq:III.1b}), must be determined through solving the
stationary form of the matter dynamics~(\ref{eq:II.20a})-(\ref{eq:II.20b}). This
stationary form (``\emph{mass eigenvalue equations''}) may be obtained either by direct
substitution of the stationary ansatz~(\ref{eq:III.1a})-(\ref{eq:III.1b}) into the coupled
Dirac equations~(\ref{eq:II.20a})-(\ref{eq:II.20b}), or by substitution of that ansatz
into the matter Lagrangean~$\LD[\Psi]$~(\ref{eq:II.58}) and then carrying out the
variational procedure with respect to the spatial parts~$\psi_a(\vec{r})$ of the wave
functions. Here it is easy to see that the matter Lagrangean~$\LD[\Psi]$ splits up into a
sum of the single-particle contributions~$\LD(a)$, i.e.
\begin{equation}
  \label{eq:III.3}
  \LD[\Psi] = \LD(1) + \LD(2)\ ,
\end{equation}
with the individual contributions being given by
\begin{subequations}
  \begin{align}
    \label{eq:III.4a}
    \LD(1) &= -i\hbar c\bar{\psi}_1\gamma^\mu(D_\mu\psi_1) - \Mp c^2\bar{\psi}_1\psi_1\\*
    \label{eq:III.4b}
    \LD(2) &= i\hbar c\bar{\psi}_2\gamma^\mu(D_\mu\psi_2) - \Me c^2\bar{\psi}_2\psi_2\ .
  \end{align}
\end{subequations}

However, observe here that these matter contributions to the Lagrangean $\LRST$ do
nevertheless contain the gauge fields which invade the matter Lagrangean via the
gauge-covariant derivatives~$D_\mu\psi_a$~(\ref{eq:II.8a})-(\ref{eq:II.8b}). Therefore the
matter contributions themselves split up into the proper kinetic parts~$\LDk(a)$ and the
electric plus magnetic interaction contributions in the following way:
\begin{equation}
  \label{eq:III.5}
  \LD(a) = \LDk(a) + \LDe(a) + \LDm(a) + \LDM(a)\ ,
\end{equation}
i.e.\ for the first (positively charged) particle
\begin{subequations}
  \begin{align}
    \label{eq:III.6a}
    \LDk_{(1)} &= -i\hbar c\,\bar{\psi}_1(\vec{r})\vec{\gamma}\sdot\vec{\nabla}\psi_1(\vec{r})\\*
    \label{eq:III.6b}
    \LDe_{(1)} &= -\hbar c\; \iiAn\cdot \ikn \\*
    \label{eq:III.6c}
    \LDm_{(1)} &= \phantom{-}\hbar c\vec{A}_2(\vec{r})\sdot\vec{k}_1(\vec{r})\\*
    \label{eq:III.6d}
    \LDM_{(1)} &= -M_1 c^2\cdot\ikn - \Mp c^2\cdot\bar{\psi}_1(\vec{r})\psi_1(\vec{r})
  \end{align}
\end{subequations}
and similarly for the second (negatively charged) particle
\begin{subequations}
  \begin{align}
    \label{eq:III.7a}
    \LDk_{(2)} &= \phantom{-}i\hbar c\,\bar{\psi}_2(\vec{r})\vec{\gamma}\sdot\vec{\nabla}\psi_2(\vec{r})\\*
    \label{eq:III.7b}
    \LDe_{(2)} &= \phantom{-}\hbar c\; \iAn\cdot \iikn \\*
    \label{eq:III.7c}
    \LDm_{(2)} &= -\hbar c\,\vec{A}_1(\vec{r})\sdot\vec{k}_2(\vec{r})\\*
    \label{eq:III.7d}
    \LDM_{(2)} &= \phantom{-}M_2 c^2\cdot\iikn - \Me c^2\cdot\bar{\psi}_2(\vec{r})\psi_2(\vec{r})\ .
  \end{align}
\end{subequations}
From this reason, the space part~($\oWD$, say)
\begin{equation}
  \label{eq:III.8}
  \oWD(a) \doteqdot \int d^3\vec{r}\; \LD(a)
\end{equation}
of the matter contribution to the action integral~$\WRST$~(\ref{eq:II.56b}) is not only
built up by the kinetic and rest mass terms but contains also the electric~(e) and
magnetic~(m) mass equivalents~$M^{(\mathrm{e,m})}_\mathrm{I/II}c^2$ of the gauge field
energy, i.e.\ for the first particle
\begin{equation}
  \label{eq:III.9}
  \oWD(1) = -\Z_{(1)}^2\cdot\Mp c^2 - M_1 c^2\cdot\int d^3\vec{r}\;\; \ikn + \int
  d^3\vec{r}\;\LDk_{(1)} -\Mei c^2 - \Mmi c^2
\end{equation}
and similarly for the second particle
\begin{equation}
  \label{eq:III.10}
  \oWD(2) = -\Z_{(2)}^2\cdot\Me c^2 + M_2 c^2\cdot\int d^3\vec{r}\;\; \iikn + \int
  d^3\vec{r}\;\LDk_{(2)} -\Meii c^2 - \Mmii c^2\ .
\end{equation}
Here the \emph{mass renormalization factors}~$\Z_{(a)}^2$ are defined through~$(a=1,2)$
\begin{equation}
  \label{eq:III.11}
  \Z^2_{(a)} = \int d^3\vec{r}\; \bar{\psi}_a(\vec{r})\psi_a(\vec{r})\ .
\end{equation}
Furthermore the electric mass equivalents of the interaction energy appear as
\begin{subequations}
  \begin{align}
    \label{eq:III.12a}
    \Mei c^2 &= \hbar c\int d^3\vec{r}\;\;\iiAn\cdot \ikn\\*
    \label{eq:III.12b}
    \Meii c^2 &= -\hbar c\int d^3\vec{r}\;\;\iAn\cdot \iikn\ ,
  \end{align}
\end{subequations}
and analogously for the magnetic mass equivalents
\begin{subequations}
  \begin{align}
    \label{eq:III.13a}
    \Mmi c^2 &= -\hbar c\int d^3\vec{r}\; \vec{A}_2(\vec{r})\sdot \vec{k}_1(\vec{r})\\*
    \label{eq:III.13b}
    \Mmii c^2 &= \hbar c\int d^3\vec{r}\; \vec{A}_1(\vec{r})\sdot
    \vec{k}_2(\vec{r})\ .
  \end{align}
\end{subequations}

But with this explicit structure of the matter Lagrangean~$\LD[\Psi]$~(\ref{eq:III.3}) it
is easy to see that the variation of the action
integral~$\WRST$~(\ref{eq:II.56a})-(\ref{eq:II.56b}) with respect to the first wave
function~$\psi_1$
\begin{equation}
  \label{eq:III.14}
  \delta_{(1)}\WRST = \delta_{(1)}\oWD{}_{(1)} \mustbe 0
\end{equation}
yields just the first mass eigenvalue equation
\begin{equation}
  \label{eq:III.15}
  i\vec{\gamma}\sdot\vec{\nabla}\psi_1(\vec{r}) + \iiAn\gamma^0\psi_1(\vec{r}) -
  \vec{A}_2(\vec{r})\sdot \vec{\gamma}\psi_1(\vec{r}) = -\left(\frac{M_1 c}{\hbar}\, \gamma_0
  + \frac{\Mp c}{\hbar}\right)\cdot\psi_1(\vec{r})\ ,
\end{equation}
and similarly the variation of $\WRST$ with respect to the second wave function~$\psi_2$
\begin{equation}
  \label{eq:III.16}
  \delta_{(2)}\WRST = \delta_{(2)}\oWD{}_{(2)} \mustbe 0
\end{equation}
yields the second mass eigenvalue equation
\begin{equation}
  \label{eq:III.17}
  i\vec{\gamma}\sdot\vec{\nabla}\psi_2(\vec{r}) + \iAn\gamma^0\psi_2(\vec{r}) -
  \vec{A}_1(\vec{r})\sdot \vec{\gamma}\psi_2(\vec{r}) = -\left(\frac{M_2 c}{\hbar}\, \gamma_0
  - \frac{\Me c}{\hbar}\right)\cdot\psi_2(\vec{r})\ .
\end{equation}

\begin{center}
  \emph{\textbf{B.\ Mass Functional}}
\end{center}

An interesting property of the mass eigenvalue equations (\ref{eq:III.15})
and (\ref{eq:III.17}) refers to the fact that they are linear with respect to the wave
functions~$\psi_a$. This implies that one can multiply any solution~$\psi_a(\vec{r})$ by
some constant and then obtains a further solution, which necessarily can not modify the
value of the matter functionals~$\oWD(a)$ upon the whole class of such solutions. On the
other hand, the matter functionals~$\oWD(a)$~(\ref{eq:III.8}) are bilinear with respect to
the wave functions and their (pseudo-) Hermitian conjugates~$\bar{\psi}_a$; and from this
one concludes that~$\oWD(a)$ must take the value zero upon the solutions of the mass eigenvalue
equations~$(a=1,2)$:
\begin{equation}
  \label{eq:III.18}
  \oWD(a) = 0\ .
\end{equation}
This is the reason why one is not forced to impose some normalization condition upon the
wave functions~$\psi_a(\vec{r})$ when one deduces the mass eigenvalue equations as the
Euler-Lagrange equations due to the RST variational
principle~(\ref{eq:II.56a})-(\ref{eq:II.56b}).

However, it is just this property~(\ref{eq:III.18}) of the matter functionals~$\oWD(a)$
which enables one to resolve these equations for the mass eigenvalues~$M_a$, where it is
convenient to apply the normalization conditions~(\ref{eq:II.44}) for the wave
functions~$\psi_a(\vec{r})$; and these \emph{mass functionals} (~$M_{[a]}c^2$, say) emerge
then in the following form:
\begin{subequations}
  \begin{align}
    \label{eq:III.19a}
    -M_{[1]} c^2 &= \Z_{(1)}^2\cdot\Mp c^2 + 2\Tkin_{(1)}+\Mei c^2 + \Mmi c^2\\*
    \label{eq:III.19b}
    \phantom{-}M_{[2]} c^2 &= \Z_{(2)}^2\cdot\Me c^2 + 2\Tkin_{(2)}+\Meii c^2 + \Mmii c^2\ .
  \end{align}
\end{subequations}
Here the mass renormalization factors~$\Z_{(a)}^2$ have already been specified by
equation~(\ref{eq:III.11}) and the mass equivalents of the interaction energies
by (\ref{eq:III.12a})-(\ref{eq:III.13b}). The emergence of a pre-factor of two in front of
the kinetic energies~$\Tkin(a)$, being given by
\begin{subequations}
  \begin{align}
    \label{eq:III.20a}
    \Tkin_{(1)} &= \phantom{-}\frac{i}{2}\hbar c\int d^3\vec{r}\;\bar{\psi_1}\vec{\gamma}\sdot
    \vec{\nabla} \psi_1(\vec{r})\\*
    \label{eq:III.20b}
    \Tkin_{(2)} &= -\frac{i}{2}\hbar c\int d^3\vec{r}\;\bar{\psi_2}\vec{\gamma}\sdot
    \vec{\nabla} \psi_2(\vec{r})\ ,
  \end{align}
\end{subequations}
is a relativistic effect and is compensated by the mass renormalization factors~$\Z_{(a)}^2$
(see the discussion of this effect in ref.s~\cite{BMS,PS}). But clearly, in place of resolving
the equations~(\ref{eq:III.18}) for the mass eigenvalues~$M_a$ in order to obtain the mass
functionals~$M_{[a]}c^2$~(\ref{eq:III.19a})-(\ref{eq:III.19b}), one could equally well
multiply through the original mass eigenvalue equations~(\ref{eq:III.15})
and~(\ref{eq:III.17}) by~$\bar{\psi}_1$ and~$\bar{\psi}_2$ and integrating over whole
three-space in order to arrive again at the same mass functionals $M_{[a]}c^2$
(\ref{eq:III.19a})-(\ref{eq:III.19b}).

The relevance of the mass functionals originates now from the fact that they are
stationary upon the solutions of the mass eigenvalue equations~(\ref{eq:III.15})
and~(\ref{eq:III.17}), albeit with regard of the constraints of wave function
normalization~(\ref{eq:II.44}). In order to take account of these constraints, one
introduces the Lagrangean multipliers~$\lD(a)$ and considers the modified mass
functionals~$\tilde{M}_{[a]}c^2$
\begin{subequations}
  \begin{align}
    \label{eq:III.21a}
    -\tilde{M}_{[1]}c^2 &= -M_{[1]}c^2 + \lD_{(1)}\cdot\ND_{(1)}\\*
    \label{eq:III.21b}
    \tilde{M}_{[2]}c^2 &= M_{[2]}c^2 + \lD_{(2)}\cdot\ND_{(2)}\ ,
  \end{align}
\end{subequations}
with the constraints~$\ND(a)$ being given by the wave function normalization~$(a=1,2)$,
i.e. 
\begin{equation}
  \label{eq:III.22}
  \ND(a) \doteqdot \int d^3\vec{r}\;\bar{\psi}_a\gamma^0\psi_a(\vec{r})-1 = 0\ .
\end{equation}
Indeed, carrying through here the variational procedure and comparing the emerging
variational equations to the former mass eigenvalue equations~(\ref{eq:III.15})
and~(\ref{eq:III.17}) just yields the identity of both sets of equations, provided one fixes
the Lagrangean multipliers~$\lD(a)$ in terms of the mass eigenvalues~$M_a$ as follows:
\begin{subequations}
  \begin{align}
    \label{eq:III.23a}
    \lD_{(1)} &= M_1 c^2\\*
    \label{eq:III.23b}
    \lD_{(2)} &= -M_2 c^2\ .
  \end{align}
\end{subequations}
Thus the Lagrangean multipliers just turn out to be identical to the mass eigenvalues (up
to sign). Clearly such a result strongly remembers one of the well-known \emph{Ritz
  variational principle} of conventional quantum mechanics which has frequently been
applied in the early days of atomic physics~\cite{BeSa} and in the meantime has been
advanced to a standard exercise for any student~\cite{Fl}.

Indeed, anyone of the mass functionals~$M_{[a]}c^2$~(\ref{eq:III.19a})-(\ref{eq:III.19b})
appears to be of a very plausible form, namely as the sum of three kinds of energy: rest
mass energy, kinetic energy and interaction energy of the electric~(e) and magnetic~(m)
type. Therefore it is very tempting to think that the total energy of the bound
two-particle system could be identified with the sum~($\tilde{\MT}c^2$) of both mass
eigenvalues
\begin{equation}
  \label{eq:III.24}
  \tilde{\MT}c^2 = -\tilde{M}_{[1]}c^2 + \tilde{M}_{[2]}c^2\ ,
\end{equation}
especially because the corresponding variational equations just coincide with the mass
eigenvalue equations~(\ref{eq:III.15}) and~(\ref{eq:III.17}). However, such a supposition
is incorrect; and a closer inspection of how the Ritz variational principle emerges as the
non-relativistic limit of the present mass functional approach will reveal the origin of
the deficiencies connected with the non-relativistic Ritz method. Furthermore this
analysis provides one with a valuable hint on how to overcome the shortcomings of
those non-relativistic variational methods and to construct the relativistic energy
functional~$\tilde{\ET}$.

\begin{center}
  \emph{\textbf{C.\ Ritz Variational Principle}}
\end{center}

In order to deduce the non-relativistic limit form of both the mass
functionals~$M_{[a]}c^2$~(\ref{eq:III.19a})-(\ref{eq:III.19b}) and of their variational
equations ~(\ref{eq:III.15}) and~(\ref{eq:III.17}) one conceives the Dirac
four-spinors~$\psi_a(\vec{r})$ as a direct sum of two-component Pauli spinors~$\ap\
(a=1,2)$, i.e.\ one puts
\begin{equation}
  \label{eq:III.25}
  \psi_a(\vec{r})=\app\oplus\apm\ ,
\end{equation}
and then one deduces the corresponding eigenvalue equations for these Pauli spinors from
the original mass eigenvalue equations~(\ref{eq:III.15}) and~(\ref{eq:III.17}). This
yields for the first particle~$(a=1)$~\cite{PS}
\begin{subequations}
  \begin{align}
    \label{eq:III.26a}
    i\vec{\sigma}\sdot\vec{\nabla}\ipp + \iiAn\cdot\ipm - \vec{A}_2 \sdot
    \vec{\sigma}\ipp &= \frac{\Mp-M_1}{\hbar}\,c\cdot\ipm\\*
    \label{eq:III.26b}
    i\vec{\sigma}\sdot\vec{\nabla}\ipm + \iiAn\cdot\ipp - \vec{A}_2 \sdot
    \vec{\sigma}\ipm &= -\frac{\Mp+M_1}{\hbar}\,c\cdot\ipp\ ,
  \end{align}
\end{subequations}
and similarly for the second particle~$(a=2)$
\begin{subequations}
  \begin{align}
    \label{eq:III.27a}
    i\vec{\sigma}\sdot\vec{\nabla}\iipp + \iAn\cdot\iipm - \vec{A}_1 \sdot
    \vec{\sigma}\iipp &= -\frac{\Me+M_2}{\hbar}\,c\cdot\iipm\\*
    \label{eq:III.27b}
    i\vec{\sigma}\sdot\vec{\nabla}\iipm + \iAn\cdot\iipp - \vec{A}_1 \sdot
    \vec{\sigma}\iipm &= \frac{\Me-M_2}{\hbar}\,c\cdot\iipp\ .
  \end{align}
\end{subequations}

Next one \emph{approximately} resolves the eigenvalue equations (\ref{eq:III.26a})
and (\ref{eq:III.27a}) for the ``small'' components~$\apm$
\begin{subequations}
  \begin{align}
    \label{eq:III.28a}
    \ipm &\backsimeq \phantom{-}\frac{i\hbar}{2\Mp c}\,\vec{\sigma}\sdot\vec{\nabla}\ipp\\*
    \label{eq:III.28b}
    \iipm &\backsimeq -\frac{i\hbar}{2\Me c}\,\vec{\sigma}\sdot\vec{\nabla}\iipp
  \end{align}
\end{subequations}
and substitutes this into the eigenvalue equations~(\ref{eq:III.26b})
and~(\ref{eq:III.27b}) for the ``small'' components~$\apm$ in order to obtain the
well-known Pauli equations for the ``large'' components:
\begin{subequations}
  \begin{align}
    \label{eq:III.29a}
    -\frac{\hbar^2}{2\Mp}\Delta\ipp + \hbar c\iiAn\cdot\ipp &= \ES_{(1)}\cdot\ipp\\*
    \label{eq:III.29b}
    -\frac{\hbar^2}{2\Me}\Delta\iipp - \hbar c\iAn\cdot\iipp &= \ES_{(2)}\cdot\iipp\ .
  \end{align}
\end{subequations}
Here the Pauli-Schr\"odinger energy eigenvalues~$\ES(a)$ are introduced through
\begin{subequations}
  \begin{align}
    \label{eq:III.30a}
    \ES_{(1)} &= -\left(\Mp c^2 + M_1c^2\right)\\*
    \label{eq:III.30b}
    \ES_{(2)} &=  M_2c^2 - \Me c^2\ ,
  \end{align}
\end{subequations}
which again demonstrates that the mass eigenvalue of the first particle is negative~$(M_1 <
0)$, in contrast to the second eigenvalue~$(M_2>0)$. Furthermore the magnetic interaction
terms~($\sim \vec{A}_a$) are omitted since the corresponding magnetic interaction energy
is mostly much smaller than its electric counterpart described by~$\aAn$. As a consequence
of this omission, the Pauli spinors can be assumed to occupy a fixed direction in spinor
space, e.g.
\begin{equation}
  \label{eq:III.31}
  \app = \varphi_a(\vec{r})\cdot|\uparrow>\ ,
\end{equation}
so that for the scalar wave functions~$\varphi_a(\vec{r})$ there emerge the conventional
Schr\"odinger equations from the Pauli system~(\ref{eq:III.29a})-(\ref{eq:III.29b}):
\begin{subequations}
  \begin{align}
    \label{eq:III.32a}
    -\frac{\hbar^2}{2\Mp}\Delta\varphi_1(\vec{r}) + \hbar c\iiAn\cdot\varphi_1(\vec{r})
    &= \ES_{(1)}\cdot\varphi_1(\vec{r})\\*
    \label{eq:III.32b}
    -\frac{\hbar^2}{2\Me}\Delta\varphi_2(\vec{r}) - \hbar c\iAn\cdot\varphi_2(\vec{r})
    &= \ES_{(2)}\cdot\varphi_2(\vec{r})\ .
  \end{align}
\end{subequations}
Observe here that either of the two particles feels the attractive force due to the other
one because the first potential~$\iAn$ (due to the positively charged particle) is
positive and the second potential~$\iiAn$ is negative! (See below for the discussion of
the corresponding Poisson equations~(\ref{eq:III.47a})-(\ref{eq:III.47d})).

Now the interesting point with this non-relativistic
approximation~(\ref{eq:III.32a})-(\ref{eq:III.32b}) to the properly relativistic
eigenvalue equations~(\ref{eq:III.15}) and~(\ref{eq:III.17}) refers to the fact that the
non-relativistic system may be also deduced from a variational principle; i.e.\ the
well-known Ritz principle~\cite{TDL}
\begin{subequations}
  \begin{align}
    \label{eq:III.33a}
    \delta \WS &= 0\\*
    \label{eq:III.33b}
    \WS &= \iint d^3\vec{r}_1\;d^3\vec{r}_2\; \sPh \hHS
    \Ph\ ,
  \end{align}
\end{subequations}
where the conventional Hamiltonian~$\hHS$ is given by
\begin{equation}
  \label{eq:III.34}
  \hHS = \frac{\vec{p}_1\,^2}{2\Mp} +  \frac{\vec{p}_2\,^2}{2\Me} - \frac{e^2}{||\vec{r}_1-\vec{r}_2||}\ .      
\end{equation}
Indeed, carrying out the variational procedure~(\ref{eq:III.33a}), with the constraint of
wave normalization, lets emerge the conventional Schr\"odinger equation as the
corresponding variational equation
\begin{equation}
  \label{eq:III.35}
  \hHS \Ph = \ES\cdot\Ph\ .
\end{equation}
It is generally believed that the associated conventional eigenvalue~$\ES$ is the
``true'' (albeit non-relativistic) energy of such a two-particle system being specified by
the Schr\"odinger Hamiltonian~(\ref{eq:III.34}); and consequently all other predictions
can at most appear to be approximations to that true value! For instance, one may try (for the
groundstate) the following product ansatz (Hartree approximation)
\begin{equation}
  \label{eq:III.36}
  \Ph = \varphi_1(\vec{r}_1)\cdot \varphi_2(\vec{r}_2)
\end{equation}
and use this for carrying out the variational
procedure~(\ref{eq:III.33a})-(\ref{eq:III.33b}) which then yields the following
one-particle eigenvalue equations:
\begin{subequations}
  \begin{align}
    \label{eq:III.37a}
    -\frac{\hbar^2}{2\Mp}\Delta\varphi_1(\vec{r}) + \hbar c\;\iiVHS\cdot\varphi_1(\vec{r})
    &= -\lS_{(1)}\cdot\varphi_1(\vec{r})\\*
    \label{eq:III.37b}
    -\frac{\hbar^2}{2\Me}\Delta\varphi_2(\vec{r}) - \hbar c\;\iVHS\cdot\varphi_2(\vec{r})
    &= -\lS_{(2)}\cdot\varphi_2(\vec{r})\ . 
  \end{align}
\end{subequations}

Here the normalization conditions for the conventional wave functions~$\varphi_a(\vec{r})$
\begin{equation}
  \label{eq:III.38}
  \int d^3\vec{r}\;\sph_a(\vec{r}) \varphi_a(\vec{r})-1=0
\end{equation}
are respected by application of the method of Lagrangean multipliers; and the
\emph{Hartree-Schr\"odinger potentials}~$\aVHS$ are given in terms of the one-particle
wave functions~$\varphi_a(\vec{r})$ through
\begin{subequations}
  \begin{align}
    \label{eq:III.39a}
    \iVHS &= \phantom{-}\as\int d^3\vec{r}_1\; \frac{\sphi\cdot\varphi_1(\vec{r}_1)}{||\vec{r}-\vec{r}_1||}\\*
    \label{eq:III.39b}
    \iiVHS &= -\as\int d^3\vec{r}_2\;\frac{\sphii\cdot\varphi_2(\vec{r}_2)}{||\vec{r}-\vec{r}_2||}\ .
  \end{align}
\end{subequations}

Clearly, the Lagrangean multipliers~$\lS(a)$ in the Hartree-Schr\"odinger eigenvalue
equations~(\ref{eq:III.37a})-(\ref{eq:III.37b}) are to be identified again with the energy
eigenvalues~$\ES(a)\ (a=1,2)$
\begin{equation}
  \label{eq:III.40}
  \lS(a) = - \ES(a)\ ,
\end{equation}
cf.\ the relativistic version~(\ref{eq:III.23a})-(\ref{eq:III.23b}); and thus \emph{the
  Hartree-Schr\"odinger system}~(\ref{eq:III.37a})-(\ref{eq:III.37b}) \emph{is identical to the
  non-relativistic RST limit (\ref{eq:III.32a})-(\ref{eq:III.32b})}, provided one can show
(see below) that the electric RST potentials~$\aAn$ are identical to the
Hartree-Schr\"odinger potentials $\aVHS$ (\ref{eq:III.39a})-(\ref{eq:III.39b}). This
important identification of the Hartree-Schr\"odinger (or more generally: Hartree-Fock)
approach with the non-relativistic limit of the RST eigenvalue equations suggests that the
RST variational method
\begin{equation}
  \label{eq:III.41}
  \delta\tilde{M}_{\mathrm{T}} = 0
\end{equation}
(with the mass functional~$\tilde{M}_{\mathrm{T}}$ given by
equations~(\ref{eq:III.21a})-(\ref{eq:III.24})) may be considered as a viable relativistic
generalization of the Hartree-Ritz variational principle (or Hartree-Fock approach, resp). But
since the latter approach is in general used as merely an \emph{approximation} to the
conventional Schr\"odinger theory, being based upon the \emph{``exact''} non-relativistic
eigenvalue equation~(\ref{eq:III.35}), it may seem that RST is afflicted with all those
deficiencies of the Hartree-Fock approach, albeit on a relativistic level (for a critical
discussion of the limits of the HF approach see ref.~\cite{Ba}). However this conclusion
is not valid because RST is capable of overcoming the main deficiencies of both the Ritz
principle and the HF approach (apart from their non-relativistic character). These
deficiencies are the following:
\begin{itemize}
\item[\textbf{i)}] the Ritz principle (\ref{eq:III.33a})-(\ref{eq:III.34}) relies upon the
  \emph{instantaneous Cou\-lomb interaction} of the particles and thus violates the true
  spirit of relativity
\item[\textbf{ii)}] the Hartree-Fock approach relies upon \emph{unique} wave
  functions~$\varphi_a(\vec{r})$ (\ref{eq:III.36}), whereas not the wave functions
  themselves (as \emph{unobservable} objects) need be unique but rather the
  \emph{observable} quantities (i.e.\ the physical densities in RST).
\end{itemize}
In the following, we will explicitly demonstrate that the treatment of the interaction
fields as proper dynamical variables together with the use of non-unique wave functions
may actually lead to predictions which can compete with those of the conventional quantum
theory. Indeed, this improvement of the standard Ritz-Hartree-Fock approaches will then
result in the desired \emph{principle of minimal energy} for the stationary bound states.
\pagebreak

\begin{center}
  \emph{\textbf{D.\ Poisson Equations}}
\end{center}

Surely, if the particle interaction is to be considered as a dynamical object, then it
must obey some equation of motion and cannot be specified by the rigid Coulomb
interaction. However in RST, the existence of such a (relativistic) equation of motion for
the interaction fields is a matter of course, since the RST variational
principle~(\ref{eq:II.56a})-(\ref{eq:II.56b}) includes the gauge field~$\A_\mu$ a priori.
It is merely necessary to carry out the variational procedure~(\ref{eq:II.56a}) with
respect to the bundle connection~$\A_\mu$ which then yields the general Maxwell
equations~(\ref{eq:II.24}), or their Abelian
specialization~(\ref{eq:II.27a})-(\ref{eq:II.27b}), resp. Since for the present Abelian
situation (for \emph{non-identical} particles) the field strengths~${F^a}_{\mu\nu}$ degenerate to
the simple curls of the potential~${A^a}_\mu$
\begin{equation}
  \label{eq:III.42}
  {F^a}_{\mu\nu} = \partial_\mu{A^a}_\nu - \partial_\nu{A^a}_\mu
\end{equation}
(see ref.~\cite{BeSo2} for the case of identical particles) the Abelian Maxwell
equations~(\ref{eq:II.27a})-(\ref{eq:II.27b}) yield just the well-known wave equations of
classical electrodynamics~$(a=1,2)$:
\begin{equation}
  \label{eq:III.43}
  \partial^\mu\partial_\mu{A^a}_\nu = 4\pi\as{j^a}_\nu\ ,
\end{equation}
provided one applies the usual Lorentz gauge condition
\begin{equation}
  \label{eq:III.44}
  \partial^\mu{A^a}_\mu\equiv 0\ .
\end{equation}

For the present stationary situation~(\ref{eq:III.2}), this splits up into the (electric)
time component~$(a=1,2)$
\begin{equation}
  \label{eq:III.45}
  \Delta\aAn = - 4\pi\as\ajn
\end{equation}
and (magnetic) space component
\begin{equation}
  \label{eq:III.46}
  \Delta\vec{A}_a(\vec{r}) = -4\pi\as\vec{j}_a(\vec{r})\ .
\end{equation}
Or, if the Maxwell currents~${j^a}_\mu=\{\ajn,-j_a(\vec{r})\}$ are written in terms of the
stationary Dirac currents~$k_{a\mu}$~(\ref{eq:II.45}) one has the Poisson
equations
\begin{subequations}
  \begin{align}
    \label{eq:III.47a}
    \Delta\iAn &= - 4\pi\as\ikn\\*
    \label{eq:III.47b}
    \Delta\iiAn &= 4\pi\as\iikn\\*
    \label{eq:III.47c}
    \Delta\vec{A}_1(\vec{r}) &= - 4\pi\as\vec{k}_1(\vec{r})\\*
    \label{eq:III.47d}
    \Delta\vec{A}_2(\vec{r}) &=  4\pi\as\vec{k}_2(\vec{r})\ .
  \end{align}
\end{subequations}
Observe here that the fibre submetric~$K_{ab}$~(\ref{eq:II.37}) has just the right
form~(\ref{eq:II.61}) in order to get the Poisson equations adapted to the positive and
negative charges carried by the particles! The standard solutions of the Poisson equations
are adopted to be of the usual form
\begin{subequations}
  \begin{align}
    \label{eq:III.48a}
    \iAn &= \as\int d^3\vec{r}\,'\frac{\ikns}{||\vec{r}-\vec{r}\,'||}\\*
    \label{eq:III.48b}
    \iiAn &= - \as\int d^3\vec{r}\,'\frac{\iikns}{||\vec{r}-\vec{r}\,'||}\\*
    \label{eq:III.48c}
    \vec{A}_1(\vec{r}) &= \as\int d^3\vec{r}\,'\frac{\vec{k}_1(\vec{r}\,')}{||\vec{r}-\vec{r}\,'||}\\*
    \label{eq:III.48d}
    \vec{A}_2(\vec{r}) &= - \as\int
    d^3\vec{r}\,'\frac{\vec{k}_2(\vec{r}\,')}{||\vec{r}-\vec{r}\,'||}\ .
  \end{align}
\end{subequations}
The non-relativistic limits of these formally exact solutions are obtained by substituting
herein the non-relativistic approximations for the Dirac densities of charge~$\akn$ and
current~$\vec{k}_a(\vec{r})$~\cite{BeSo2}:
\begin{subequations}
  \begin{align}
    \label{eq:III.49a}
    \akn &= \bar{\psi}_a(\vec{r})\gamma_0\psi_a(\vec{r}) = \sapp\sdot\app + \sapm\sdot\apm\\*\notag
    &\backsimeq \overset{*}{\varphi}_a(\vec{r})\cdot\varphi_a(\vec{r})\\*
    \label{eq:III.49b}
    \vec{k}_a(\vec{r}) &= \bar{\psi}_a(\vec{r})\sdot\vec{\gamma}\sdot\psi_a(\vec{r}) =
    \sapp\sdot\vec{\sigma}\sdot\apm + \sapm\sdot \vec{\sigma}\sdot\app
  \end{align}
\end{subequations}

Observe here that for the non-relativistic limit the ``small'' Pauli components~$\apm$ can
be neglected against its ``large'' counterparts~$\app$ only for the charge
density~$\akn$~(\ref{eq:III.49a}), but not for the current densities~(\ref{eq:III.49b}) as
the sources of the magnetic fields! This is the reason why we omitted the magnetic terms
when deducing the Hartree-Schr\"odinger eigenvalue
equations~(\ref{eq:III.32a})-(\ref{eq:III.32b}) for the ``large'' components from their
properly relativistic RST form~(\ref{eq:III.15}) and~(\ref{eq:III.17}). But inserting now
the approximate form~(\ref{eq:III.49a}) of the charge densities~$\akn$ into the formal
solutions~$\aAn$~(\ref{eq:III.48a})-(\ref{eq:III.48b}) of the Poisson equations yields
\begin{subequations}
  \begin{align}
    \label{eq:III.50a}
    \iAn &\backsimeq \as\int d^3\vec{r}\,
    '\;\frac{\sphis\cdot\varphi_1(\vec{r}\,')}
    {||\vec{r}-\vec{r}\,'||}\\*
    \label{eq:III.50b}
    \iiAn &\backsimeq -\as\int
    d^3\vec{r}\,'\;\frac{\sphiis\cdot\varphi_2(\vec{r}\,')}{||\vec{r}-\vec{r}\,'||}\ ,
  \end{align}
\end{subequations}
and this is just the result which is needed in order to identify the non-relativistic RST
limit~(\ref{eq:III.32a})-(\ref{eq:III.32b}) with the conventional
Hartree-Schr\"odinger eigenvalue system (\ref{eq:III.37a})-(\ref{eq:III.37b}).

This is the way in which RST cures the first one~\textbf{(i)} of the Ritz-Hartree-Fock deficiencies
mentioned above, namely by adopting the Poisson equations (or more generally: the Maxwell
equations) for the determination of the gauge potentials and thus accepting the action of
the gauge forces as a truly dynamical process. However, observe in this context that the
Poisson equations emerge here as the variational equations due to the Hamilton-Lagrange
action principle~(\ref{eq:II.56a})-(\ref{eq:II.56b}), not due to the variation of the RST
mass functional~$\tilde{M}_T c^2$~(\ref{eq:III.24})! Therefore it is not yet possible to
identify this mass functional with the wanted energy functional as the ultimate goal of
the present investigation. This goal will be readily achieved in the next section but can
be prepared here by first regarding an important property of the time-independent gauge
part~$\oWG[\A_\mu]$ 
\begin{equation}
  \label{eq:III.51}
  \oWG[\A_\mu] = \int d^3\vec{r}\;\LG[\A_\mu]
\end{equation}
of the original RST action principle~(\ref{eq:II.56b}). Indeed, substituting here the
Lagrangean density~$\LG$ in the form~(\ref{eq:II.62}) with the electromagnetic
three-vector fields~$\vec{E}_a(\vec{r})$ and~$\vec{H}_a(\vec{r})$ being written in terms
of the corresponding potentials~$\aAn,\vec{A}_a(\vec{r})$ as
\begin{subequations}
  \begin{align}
    \label{eq:III.52a}
    \vec{E}_a(\vec{r}) &= - \vec{\nabla}\aAn\\*
    \label{eq:III.52b}
    \vec{H}_a(\vec{r}) &= \vec{\nabla}\times\vec{A}_a(\vec{r})
  \end{align}
\end{subequations}
lets appear that time-independent gauge part~$\oWG[\A_\mu]$ in the following form:
\begin{equation}
  \label{eq:III.53}
   \oWG[\A_\mu] = \frac{\hbar c}{4\pi\as}\int d^3\vec{r}\left\{\vec{\nabla}\,\iAn
       \sdot \vec{\nabla}\,\iiAn - \left(\vec{\nabla}\times\vec{A}_1(\vec{r}) \right)\sdot
       \left(\vec{\nabla}\times\vec{A}_2(\vec{r}) \right) \right\}
\end{equation}
Furthermore, by resorting to the gauge field contributions $\LD^{(\mathrm{e,m})}$
(\ref{eq:III.6b}), (\ref{eq:III.6c}) and (\ref{eq:III.7b}), (\ref{eq:III.7c}) due to the
matter Lagrangean~$\LD$, one has the corresponding electric~(e) and magnetic~(m) action
constituents as
\begin{subequations}
  \begin{align}
    \label{eq:III.54a}
    \oWD^\mathrm{(e)} &= -\hbar c\int d^3\vec{r}\;\left\{\iiAn\cdot\ikn - \iAn\cdot\iikn \right\}\\*
    \label{eq:III.54b}
    \oWD^\mathrm{(m)} &= \phantom{-}\hbar c\int d^3\vec{r}\;\left\{\vec{A}_2(\vec{r})\sdot\vec{k}_1(\vec{r}) 
    - \vec{A}_1(\vec{r})\sdot\vec{k}_2(\vec{r}) \right\}\ .
  \end{align}
\end{subequations}
Thus the former Poisson equations~(\ref{eq:III.47a})-(\ref{eq:III.47d}) are actually
recovered from here by variation of the partial sum of action
integrals~$\oWG+\oWD^\mathrm{(e)}+\oWD^\mathrm{(m)}$ with respect to the static gauge
potentials.

Now the interesting point with this variational procedure for the gauge fields is that it
leads us to global identities which subsequently will be needed as constraints for the
principle of minimal energy, i.e.\ the \emph{Poisson identities}. These global
relations between the gauge fields and their  sources emerge from the Hamiltonian-Lagrange
action principle~(\ref{eq:II.56a})-(\ref{eq:II.56b}) by considering the scaling variations
for the potentials, e.g.\ for the first electrostatic potential
\begin{equation}
  \label{eq:III.55}
  \iAn \rightarrow {}^{(1)}\!\!A'_0(\vec{r}) = C_*\cdot\iAn\ ,
\end{equation}
with the scaling factor~$C_*$ being a constant over three-space. Similar arguments do hold
also for the other potentials~$\iiAn,\vec{A}_a(\vec{r})$. By inserting this special
variation~(\ref{eq:III.55}) into the RST action integral, its relevant parts become
  \begin{multline}
  \label{eq:III.56}
    \oWG^\mathrm{(e)} + \oWD^\mathrm{(e)} \Rightarrow \\*
    C_*\left[\frac{\hbar c}{4\pi\as}
    \int d^3\vec{r}\;\left(\vec{\nabla}\,\iAn\cdot\vec{\nabla}\,\iiAn + 4\pi\as\iAn\cdot\iikn
    \right) \right]\ .    
  \end{multline}
Since the Hamiltonian-Lagrange action principle demands stationarity of the action
integral~$\WRST$ with respect to the choice of~$C_*$:
\begin{equation}
  \label{eq:III.57}
  \frac{d\WRST(C_*)}{dC_*} \Big|_{C_*=1} = 0\ ,
\end{equation}
one concludes from equation~(\ref{eq:III.56}) that the following integral relation must
hold:
\begin{equation}
  \label{eq:III.58}
  N^\mathrm{(e)}_\mathrm{G}(1) \doteqdot \int d^3\vec{r}\;\left[ \left(\vec{\nabla}
     \,\iAn\right) \sdot  \left(\vec{\nabla}\,\iiAn\right) + 4\pi\as\iAn\cdot\iikn
  \right] \equiv 0\ ,
\end{equation}
and analogously for the other gauge potentials
\begin{subequations}
  \begin{align}
    \label{eq:III.59a}
     N^\mathrm{(e)}_\mathrm{G}(2) & \doteqdot \int d^3\vec{r}\;\left[ \left(\vec{\nabla}
     \,\iAn\right) \sdot  \left(\vec{\nabla}\,\iiAn\right) - 4\pi\as\iiAn\cdot\ikn
  \right] \equiv 0\\*
  \label{eq:III.59b}
  N^\mathrm{(m)}_\mathrm{G}(1)  & \doteqdot \int d^3\vec{r}\; \left[
    \left(\vec{\nabla}\times\vec{A}_1(\vec{r}) \right)\sdot \left(\vec{\nabla}\times\vec{A}_2(\vec{r})\right) +
    4\pi\as \vec{A}_1(\vec{r}) \sdot \vec{k}_2(\vec{r}) \right]\equiv 0\\*
    \label{eq:III.59c}
  N^\mathrm{(m)}_\mathrm{G}(2)  & \doteqdot \int d^3\vec{r}\; \left[
    \left(\vec{\nabla}\times\vec{A}_1(\vec{r}) \right)\sdot \left(\vec{\nabla}\times\vec{A}_2(\vec{r})\right) -
    4\pi\as \vec{A}_2(\vec{r}) \sdot \vec{k}_1(\vec{r})   \right]\equiv 0\ .
  \end{align}
\end{subequations}

Clearly, these \emph{Poisson identities} may be obtained also directly from the Poisson
equations~(\ref{eq:III.47a})-(\ref{eq:III.47d}) by multiplying through with the
appropriate potentials and integrating by parts. But their deduction from the RST action
principle does better elucidate their meaning for the variational procedure: obviously,
when looking (by trial and error) for those gauge potentials which yield stationarity of
the action integral~$\WRST$~(\ref{eq:II.56b}), one can restrict oneself to those
potentials which obey the Poisson identities. It is just with reference to this meaning of
restrictive conditions that the Poisson identities will readily be used in order to set up
the RST principle of minimal energy!

\begin{center}
  \emph{\textbf{E.\ Double-Valued Wave Functions}}
\end{center}

After the first deficiency \textbf{(i)} of the Ritz-Hartree-Schr\"odinger approach is
now eliminated, one can turn to the next critical point, namely the conventional
assumption~\textbf{(ii)} that the wave functions must always be unique. Indeed we will
relax now this presumption and will (as a counterexample) admit double-valued wave
functions of the type
\begin{equation}
  \label{eq:III.60}
  \Psi(r,\vartheta,\phi+2\pi) = -  \Psi(r,\vartheta,\phi)
\end{equation}
where~$\{r,\vartheta,\phi\}$ are the usual spherical polar coordinates. We will readily
see that such a more general class of wave functions can generate unconventional gauge
potentials, namely via the solutions of the Poisson
equations~(\ref{eq:III.48a})-(\ref{eq:III.48d}) or their non-relativistic approximations
resp; and the corresponding unusual form of interaction force may then yield energy levels
which are closer to the conventional Schr\"odinger predictions than it is possible for the
Dirac-Fock approach~\cite{Wi}. But clearly, the admitted non-uniqueness~(\ref{eq:III.60})
of the wave functions must not imply the non-uniqueness of the physical densities, e.g.\
of four-current~$k_{a\mu}$ (\ref{eq:II.45}) or of energy-momentum
density~$\DT_{\mu\nu}$~(\ref{eq:II.50b}), etc. Indeed, it is easy to see that those
physical densities are bilinear constructions of~$\Psi$ and~$\bar{\Psi}$ and therefore
remain invariant against the change~(\ref{eq:III.60}) of the wave function!

For a concrete exemplification of those double-valued wave functions, one may resort to two
basis systems~$\{\onp,\onm\}$ and~$\{\oip,\oim\}$ of the two-dimensional Pauli spinor
space which are eigenvectors (with zero eigenvalue) of the total angular momentum
$\hat{J}_z\doteqdot \hat{L}_z+\hat{S}_z$ :
\begin{equation}
  \label{eq:III.61}
  \hat{J}_z\onp = \hat{J}_z\onm = \hat{J}_z\oip = \hat{J}_z\oim = 0\ ,
\end{equation}
see ref.~\cite{BMS} for the details. Since these basis spinors themselves are already
double-valued (i.e.~$\onp(r,\vartheta,\phi+2\pi)=-\onp(r,\vartheta,\phi)$; etc.),
one can decompose the Pauli spinors~$\ap$~(\ref{eq:III.25}) with respect to these
double-valued basis systems as follows~$(a=1,2)$
\begin{subequations}
  \begin{align}
    \label{eq:III.62a}
    \app &= \phantom{-}(r\sin\vartheta)^{-\frac{1}{2}}\left[\aRp\rt\cdot\onp+\aSp\rt\cdot\onm\right]\\*
    \label{eq:III.62b}
    \apm &= -i(r\sin\vartheta)^{-\frac{1}{2}}\left[\aRm\rt\cdot\oip+\aSm\rt\cdot\oim\right]\\*
   &\left({}^{(a)}\varphi_\pm (r,\vartheta,\phi+2\pi) = -{}^{(a)}\varphi_\pm(r,\vartheta,\phi)\right)\ ,\notag        
  \end{align}
\end{subequations}
and then both these Pauli spinors and the corresponding Dirac spinors~$\psi_a(\vec{r})$
are double-valued in the sense of equation~(\ref{eq:III.60}), provided the
wave amplitude $\aRpm\rt$ and $\aSpm\rt$ are \emph{single-valued!} Moreover, the latter
property of uniqueness is transferred  also to the Dirac
densities $k_{a\mu}(\vec{r})=\{\akn,-\vec{k}_a(\vec{r}) \}$ (\ref{eq:II.45})
since these appear as bilinear constructions of the Pauli spinors~$\ap$, i.e.
\begin{subequations}
  \begin{multline}
    \label{eq:III.63a}
    \akn = \sapp\sdot\app + \sapm\sdot\apm = \\*
    \frac{\aRp^2+\aSp^2+\aRm^2+\aSm^2}{4\pi r\sin\vartheta}
    \end{multline}
    \begin{align}
    \label{eq:III.63b}
    \vec{k}_a(\vec{r}) = \sapp\sdot\vec{\sigma}\sdot\apm +
    \sapm\sdot\vec{\sigma}\sdot\app \doteqdot {}^{(a)}k_\phi(\vec{r})\cdot\vec{e}_\phi\ ,
    \end{align}
\end{subequations}
with the azimuthal component~${}^{(a)}k_\phi$ of the Dirac
currents~$\vec{k}_a(\vec{r})$ being given by
\begin{equation}
  \label{eq:III.64}
  {}^{(a)}k_\phi = \frac{\sin\vartheta\left(\aRp\cdot\aRm-\aSp\cdot\aSm\right)
  -\cos\vartheta\left(\aSp\cdot\aRm+\aRp\cdot\aSm\right)}{2\pi r\sin\vartheta}\ .
\end{equation}
Observe here, that through the choice of \emph{real-valued} wave amplitudes\\
$\aRpm,\aSpm$, the radial~$({}^{(a)}k_r)$ and longitudinal~$({}^{(a)}k_\vartheta)$ components of
the Dirac currents~$\vec{k}_a(\vec{r})$ do vanish
(i.e.~${}^{(a)}k_r={}^{(a)}k_\vartheta\equiv 0$), so that these
three-currents~$\vec{k}_a(\vec{r})$ encircle the axis of the spherical polar
coordinates~$(r,\vartheta,\phi)$. Naturally, this symmetry of the three-currents may then be
transferred also to the vector potentials~$\vec{A}_a(\vec{r})$ which thus appear in the
following form
\begin{equation}
  \label{eq:III.65}
  \vec{A}_a(\vec{r}) = {}^{(a)}A_\phi(r,\vartheta)\cdot\vec{e}_\phi\ ,
\end{equation}
from which the magnetic fields~$\vec{H}_a(\vec{r})$ can be computed by means of the usual
curl relation~(\ref{eq:III.52b}).

Summarizing the properties of the wave functions to be used for the RST description of
bound states, one first has to mention their double-valuedness (\ref{eq:III.60}) and
moreover we will assume that the wave amplitudes~$\aRpm,\aSpm$ are unique and non-singular
(real-valued) functions over space time. But observe here that, through this second
assumption, the Pauli spinors $\ap$ (\ref{eq:III.62a})-(\ref{eq:III.62b}) and therefore
also the original Dirac spinors~$\psi_a(\vec{r})$~(\ref{eq:III.25}) become both singular
and double-valued~($\leadsto$ \emph{``exotic states''}). It should appear as a matter of
course that such exotic states will imply further unconventional elements of the theory,
e.g. the form of the gauge potentials. Notice, however, that the observable objects of the
theory (i.e.\ the densities of charge, current, energy-momentum etc.) are well-defined and
unique objects over space-time, the singularities of which (if present at all) do not
induce any pathological element into the theory. Therefore the wave
amplitudes~$\aRpm,\aSpm$~(\ref{eq:III.62a})-(\ref{eq:III.62b}) as the unique and (mostly)
regular constituents of the non-unique and singular wave functions~$\psi_a(\vec{r})$~(\ref{eq:III.25})
will appear as the solutions of a well-defined eigenvalue problem. The corresponding
eigenvalue equations are to be deduced from the original eigenvalue
equations (\ref{eq:III.26a})-(\ref{eq:III.27b}) for the double-valued Pauli spinors~$\ap$
and appear then in the following form, e.g.\ for the first particle~$(a=1)$~\cite{BeSo2}
\enlargethispage{1cm}
\begin{subequations}
  \begin{align}
    \label{eq:III.66a}
    \begin{split}
    \frac{\partial\iRp}{\partial r}+\frac{1}{r}\frac{\partial\iSp}{\partial\vartheta}
    +{}^{(2)}\!A_0\cdot\iRm-{}^{(2)}\!A_\phi\left(\sin\vartheta\cdot\iRp-\cos\vartheta\cdot\iSp\right)\\
    = \frac{\Mp-M_1}{\hbar}\,c\cdot\iRm  
    \end{split}\\
    \label{eq:III.66b}
    \begin{split}
    \frac{\partial\iSp}{\partial r}-\frac{1}{r}\frac{\partial\iRp}{\partial\vartheta}
    +{}^{(2)}\!A_0\cdot\iSm+{}^{(2)}\!A_\phi\left(\sin\vartheta\cdot\iSp+\cos\vartheta\cdot\iRp\right)\\
    = \frac{\Mp-M_1}{\hbar}\,c\cdot\iSm
    \end{split}\\
    \label{eq:III.66c}
    \begin{split}
    \frac{1}{r}\frac{\partial(r\iRm)}{\partial r}-\frac{1}{r}\frac{\partial\iSm}{\partial\vartheta}
    -{}^{(2)}\!A_0\cdot\iRp+{}^{(2)}\!A_\phi\left(\sin\vartheta\cdot\iRm-\cos\vartheta\cdot\iSm\right)\\
    = \frac{\Mp+M_1}{\hbar}\,c\cdot\iRp
    \end{split}\\
    \label{eq:III.66d}
    \begin{split}
    \frac{1}{r}\frac{\partial(r\iSm)}{\partial r}+\frac{1}{r}\frac{\partial\iRm}{\partial\vartheta}
    -{}^{(2)}\!A_0\cdot\iSp-{}^{(2)}\!A_\phi\left(\sin\vartheta\cdot\iSm+\cos\vartheta\cdot\iRm\right)\\
    = \frac{\Mp+M_1}{\hbar}\,c\cdot\iSp      
    \end{split}
  \end{align}
\end{subequations}
\\An analogous set of four eigenvalue equations does apply to the second particle~$(a=2)$
which, however, needs not explicitly be reproduced here because it can be obtained simply by
means of the particle permutation symmetry, see ref.~\cite{BeSo2}.

\begin{center}
  \emph{\textbf{F.\ Unconventional Potentials}}
\end{center}

The interesting point with these double-valued wave functions is now that they do generate
a rather unusual form of the gauge potentials~$\aAn$ and~$\vec{A}_a(\vec{r})$ by means of
the recipe~(\ref{eq:III.48a})-(\ref{eq:III.48d}). In order to see this more clearly, one
substitutes the charge and current densities~$\ikn$~(\ref{eq:III.63a})
and~$\vec{k}_a(\vec{r})$~(\ref{eq:III.63b}) into those formal solutions of the Poisson
equations which yields explicitly for the electric potentials in terms of the
\emph{unique} wave amplitudes
\begin{subequations}
  \begin{multline}
    \label{eq:III.67a}
    {}^{(1)}\!A_0\rt =\\*
    \phantom{-} \frac{\as}{4\pi}\int\frac{d^3\vec{r}\,'}{r'\sin\vartheta'}
    \frac{\iRp^2\rts+\iSp^2\rts+\iRm^2\rts+\iSm^2\rts}{||\vec{r}-\vec{r}\,'||}
    \end{multline}
    \begin{multline}
    \label{eq:III.67b}
    {}^{(2)}\!A_0\rt =\\*
    -\frac{\as}{4\pi}\int\frac{d^3\vec{r}\,'}{r'\sin\vartheta'}
    \frac{\iiRp^2\rts+\iiSp^2\rts+\iiRm^2\rts+\iiSm^2\rts}{||\vec{r}-\vec{r}\,'||}
    \end{multline}
\end{subequations}
and similarly for the magnetic potentials~${}^{(a)}A_\phi\rt$.

In order to estimate qualitatively the new feature of these potentials due to the non-singular
wave amplitudes~$\aRpm,\aSpm$, it may be sufficient for the moment to adopt a simple
model~${}^{(b)}k_0(\vec{r})$ for a non-spherically symmetric and singular charge
distribution being normalized to unity, cf.~(\ref{eq:II.44}), i.e.\ we put
\begin{equation}
  \label{eq:III.68}
  {}^{(b)}k_0(r) \doteqdot\frac{{}^{(b)}\tilde{k}_0(r)}{4\pi r\sin\vartheta} =
  (4\pi r\sin\vartheta)^{-1}\cdot\frac{8}{\pi r_*^2}\exp\left({\ds -2\frac{r}{r_*}}\right)
\end{equation}
with the regular and spherically symmetric charge
distribution~${}^{(b)}\tilde{k}_0(r)$ being normalized as follows:
\begin{equation}
  \label{eq:III.69}
  \int_0^\infty dr\,r\; {}^{(b)}\tilde{k}_0(r)=\frac{2}{\pi}\ .
\end{equation}
The corresponding electric potential~$\pAn$
\begin{equation}
  \label{eq:III.70}
  \pAn = \as\int \frac{d^3\vec{r}\,'}{4\pi
    r'\sin\vartheta'}\;\frac{{}^{(b)}\tilde{k}_0(\vec{r}\,')}
  {||\vec{r}-\vec{r}\,'||}
\end{equation}
will then be found to be also non-spherically symmetric, but it can be shown~\cite{BMS} that
the binding ability of this potential is supplied mainly by its spherically symmetric
part~($\ppAn$, say). The latter part may be defined by suitable expansion of the
denominator~$||\vec{r}-\vec{r}\,'||$ in the integral~(\ref{eq:III.70}), see ref.~\cite{BMS}
for this method; or otherwise one may substitute the anisotropic
density~${}^{(b)}k_0(\vec{r})$~(\ref{eq:III.68}) into the RST action
principle~(\ref{eq:II.56a})-(\ref{eq:II.56b}) and may then determine the desired isotropic
part~$\ppAn$ of~$\pAn$~(\ref{eq:III.70}) via the solution of the corresponding
variational (i.e.\ Poisson) equation. Resorting here to the second method it suffices
to consider merely the electrostatic part~$\WRST^\mathrm{(e)}$ of the two-particle action
integral~(\ref{eq:II.56b}) with ${}^{(1)}\!A_0=-{}^{(2)}\!A_0\doteqdot
{}^{[\textrm{p}]}\!A_0$ and ${}^{(1)}\tilde{k}_0={}^{(2)}\tilde{k}_0\doteqdot
{}^{(b)}\tilde{k}_0$ which yields
\begin{equation}
  \label{eq:III.71}
  \begin{split}
   \oWRSTe &= \int d^3\vec{r}\;\left(2\LD^\mathrm{(e)}+\LG^\mathrm{(e)}\right)\\*
   &=\int d^3\vec{r}\;\left(2\hbar c\; \ppAn\cdot {}^{(b)}k_0(\vec{r})
     -\frac{\hbar c}{4\pi\as}\, ||\vec{\nabla}\,\ppAn||^2 \right)\\*
   &=\hbar c  \int dr\;r\left[\pi\,\ppAn\; {}^{(b)}\tilde{k}_0(r) - \frac{r}{\as}
     \left(\frac{d\,\ppAn}{dr}\right)^2 \right]\ .
  \end{split}
\end{equation}
Thus the electrostatic variational equation~$(\delta\oWRST=0)$ emerges as a spherically
symmetric Poisson equation:
\begin{equation}
  \label{eq:III.72}
  \left(\frac{d^2}{dr^2}+\frac{2}{r}\frac{d}{dr}\right)\ppAn=-\frac{\pi}{2}\as
  \frac{{}^{(b)}\tilde{k}_0(r)}{r}\ .
\end{equation}
Finally, substituting here the assumed charge
density~${}^\mathrm{(b)}\tilde{k}_0(r)$~(\ref{eq:III.68}) yields for the electric
potential~$\ppAn$
\begin{equation}
  \label{eq:III.73}
  \ppAn = \frac{\as}{r}\left(1-\exp\left[ {\ds -2\frac{r}{r_*}}\right] \right)\ .
\end{equation}

This interaction potential, being typical for the exotic quantum states, has some peculiar
properties: First, it approaches the Coulomb potential~$(\sim \frac{\ds \as}{r})$ at
spatial infinity~($r\to\infty$) as expected since it is generated by just one electric
charge unit (i.e.\ elementary charge). Second, the potential remains finite at the
origin~$(r=0)$ 
\begin{equation}
  \label{eq:III.74}
  {}^{[\textrm{p}]}\!A_0(0)=\frac{2\as}{r_*}
\end{equation}
as well as the corresponding electric field strength~$\vec{E}_\mathrm{p}$~(\ref{eq:III.52a})
\begin{equation}
  \label{eq:III.75}
  {}^{\mathrm{[p]}}\!E_r\Big|_{r=0}=-\frac{d\,\ppAn}{dr}\Big|_{r=0}=-\frac{2\as}{r_*^2}\ .
\end{equation}
Moreover, if the length parameter~$r_*$ tends to zero~$(r_*\to 0)$, both the
potential~(\ref{eq:III.74}) and its field strength~(\ref{eq:III.75}) approach infinity
which says that the asymptotic Coulomb form fills then the whole three-space. This is
clear because, in this limit~$(r_*\to 0)$, the charge
distribution~${}^\mathrm{(b)}\tilde{k}_0(r)$ becomes pointlike (see fig.~1 of
ref.~\cite{BMS} for a sketch of the unconventional potentials).

However from the physical point of view, the most interesting feature of those
potentials~$\aAn$ due to the exotic quantum states surely refers to the fact that they
carry a \emph{finite} energy content, in contrast to the Coulomb potential. Indeed, the
electrostatic interaction energy~$\hat{E}_\mathrm{R}^\mathrm{(e)}$ of both charges is
given by~\cite{BMS}
\begin{equation}
  \label{eq:III.76}
  \heER = -\frac{\hbar c}{4\pi\as}\int d^3\vec{r}\;||\vec{\nabla}\,{}^{(p)}\!\!A_0(\vec{r}) ||^2\ ,
\end{equation}
and when the spherically symmetric approximation~$\ppAn$~(\ref{eq:III.73}) is substituted
herein, one finds the following result
\begin{equation}
  \label{eq:III.77}
   \heER = - \frac{e^2}{r_*}\ .
\end{equation}
Incidentally, this is just the interaction energy of two point charges separated by the
distance~$r_*$ which plays the role of a length parameter for our model charge
distribution~${}^\mathrm{(b)}\tilde{k}_0(r)$~(\ref{eq:III.68}). This charge distribution
becomes pointlike when the length parameter~$r_*$ tends to zero and, clearly, for this
limit the interaction energy~$\hat{E}_\mathrm{R}^\mathrm{(e)}$~(\ref{eq:III.77}) of both
extended charge distributions becomes infinite, just as is the case with two point charges
of vanishing separation~$(r_*\to 0)$.

Of course, the interaction energy~$\heER$ of the two particles is only a fraction of their
total energy~$\ET$ which must contain also the kinetic form of the particle energy.
Indeed, this latter form of energy deserves a closer inspection, too; and this can be
performed most adequately by setting up now the total energy functional~$\ET$ through
adequately exploiting the intrinsic RST logic.
%%% Local Variables: 
%%% mode: latex
%%% TeX-master: "main_minenergy"
%%% End: 

\section{Energy Functional}
\indent

Besides the use of exotic quantum states and their unconventional potentials, it is
necessary to introduce a further new element into the theory in order to deal successfully
with the energy spectra of the bound systems: This refers to the construction of a
suitable energy functional~$\ET$, which equips the RST field configurations with an energy
content being then immediately accessible to spectroscopic test. Recall here the fact
that, though the variational approach~(\ref{eq:III.41}) due to the total mass
functional~$\tMT c^2$~(\ref{eq:III.24}) can be viewed as the relativistic generalization
of the Ritz variational principle~(\ref{eq:III.33a})-(\ref{eq:III.33b}), this relativistic
approach nevertheless \emph{fails} to establish the gauge field equations and
\emph{exclusively} reproduces the eigenvalue equations for the matter fields; see the
critical comments~\textbf{(i)} and~\textbf{(ii)} mentioned above. Therefore it suggests
itself to restart from the original notion of \emph{field
  energy}~$\ET$~(\ref{eq:II.53})-(\ref{eq:II.55b}) and to convert this to the wanted
energy functional~$\tET$.

\begin{center}
  \emph{\textbf{A.\ Relativistic Construction}}
\end{center}

First, observe that the underlying energy-momentum densities~$\DT_{\mu\nu}$
and~$\GT_{\mu\nu}$ are already specified by equations~(\ref{eq:II.50a})-(\ref{eq:II.52});
and if one substitutes therein the stationary form of the matter and gauge
fields~(\ref{eq:III.1a})-(\ref{eq:III.2}), one finds the individual energy
contributions~$\ED$ and~$\EG$~(\ref{eq:II.55a})-(\ref{eq:II.55b}) appearing in the
following form~\cite{BeSo2}:
\begin{subequations}
  \begin{align}
    \label{eq:IV.1a}
    \ED &= -\left(M_1 c^2 + \Mei c^2\right) + \left(M_2 c^2 - \Meii c^2\right) = \MT c^2 -
    \left(\Meii c^2 + \Mei c^2\right)\\*
    \label{eq:IV.1b}
    \EG&\rightarrow \hat{E}_{\mathrm{R}} = \frac{\hbar c}{4\pi\as}\int d^3 \vec{r}\, \left[
      \vec{E}_1(\vec{r})\sdot\vec{E}_2(\vec{r}) + \vec{H}_1(\vec{r}) \sdot
      \vec{H}_2(\vec{r}) \right] \doteqdot \heER + \hmER\ .
  \end{align}
\end{subequations}
Here, the mass equivalents~$\Mei c^2$ and~$\Meii c^2$ of the electrostatic interaction
energy have already been defined previously through
equations~(\ref{eq:III.12a})-(\ref{eq:III.12b}). Furthermore, the gauge field energy~$\EG$
consists exclusively of the energy content~$\hat{E}_{\mathrm{R}}$ due to the real gauge field
modes~${A^a}_\mu$, since the complex field modes~$B_\mu$ must be put to zero together with
their energy content~$\hat{E}_\mathrm{C}$ (see the discussion below equation~(\ref{eq:II.26})).

Clearly it is very tempting now to consider the total energy functional
$\ET$~(\ref{eq:II.53}), with~$\ED$ and~$\EG$ being specified by the present equations
(\ref{eq:IV.1a})-(\ref{eq:IV.1b}), as the wanted object of our interest. The corresponding
variational procedure $(\delta\ET=0)$ must then be complemented by the former constraints
of wave function normalization~(\ref{eq:III.22}), which had to be applied already in
connection with the mass functional approach~(\ref{eq:III.41}). By this arrangement, one
would be led to the following first proposal~$\tETi$ for the desired energy functional:
\begin{equation}
  \label{eq:IV.2}
  \begin{split}
    \tETi &= \ET + \lD_{(1)}\cdot \ND_{(1)} + \lD_{(2)}\cdot \ND_{(2)}\\*
    &= \tMT c^2 + \left( \heER - \Mei c^2 - \Meii c^2 \right) + \hmER \ .
  \end{split}
\end{equation}
But observe here that the additional appearance of the electrostatic mass
equivalents~$\Mei$ and~$\Meii$ does spoil the partial success already obtained with the
mass functional~$\tMT c^2$~(\ref{eq:III.24}) from which the matter eigenvalue equations
can actually be deduced. The reason is that those mass
equivalents~(\ref{eq:III.12a})-(\ref{eq:III.13b}) do also contain the wave
functions~$\psi_a(\vec{r})$, namely via the
densities~${}^{(a)}k_\mu(\vec{r})$~(\ref{eq:III.49a})-(\ref{eq:III.49b}).  Therefore we
have to eliminate again these redundant mass equivalents from our first
proposal~(\ref{eq:IV.2}), which can be achieved by expressing them in terms of the
electromagnetic gauge field energy~$\heER$ and~$\hmER$~(\ref{eq:IV.1b}) as follows:
\begin{subequations}
  \begin{align}
  \label{eq:IV.3a}
  \heER &= \Mei c^2 = \Meii c^2\\*
  \label{eq:IV.3b}
  \hmER &= -\Mmi c^2 = -\Mmii c^2\ ,
  \end{align}
\end{subequations}
and these relations represent nothing else than the Poisson identities~(\ref{eq:III.58})-(\ref{eq:III.59c}).
Thus using this electric coincidence~(\ref{eq:IV.3a}) in order to eliminate the electric
mass equivalents from the previous proposal~$\tETi$~(\ref{eq:IV.2}), one arrives at the
next proposal~$\tETii$:
\begin{equation}
  \label{eq:IV.4}
  \tETii = \tMT c^2 - \heER + \hmER\ .
\end{equation}
This second proposal displays now some pleasant features and therefore must be expected to
come close to the wanted final result: First,~$\tETii$ contains the matter
fields~$\psi_a(\vec{r})$ only in form of the mass functional~$\tMT c^2$ and therefore the
variational equations of the functional~$\tETii$ must correctly reproduce the mass
eigenvalue equations~(\ref{eq:III.15}) and~(\ref{eq:III.17})! Second, returning for the
moment to the matter energy~$\ED$~(\ref{eq:IV.1a}) and substituting there the mass
eigenvalues~$M_a c^2$~(\ref{eq:III.19a})-(\ref{eq:III.19b}) lets the matter energy~$\ED$
appear essentially as a \emph{sum of single-particle contributions}
\begin{equation}
  \label{eq:IV.5}
  \ED = \sum_{a=1}^{2} \ED_{(a)}
\end{equation}
with the individual contributions~$\ED_{(a)}$ being given by
\begin{subequations}
  \begin{align}
    \label{eq:IV.6a}
    \ED{}_{(1)} &= -\left( M_{[1]} c^2 + \Mei\right) =
    \Z_{(1)}^2\cdot\Mp c^2 + 2\Tkin{}_{(1)} + \Mmi c^2\\*
    \label{eq:IV.6b}
    \ED{}_{(2)} &= \left( M_{[2]} c^2 - \Meii c^2\right) =
    \Z_{(2)}^2\cdot\Me c^2 + 2\Tkin{}_{(2)} + \Mmii c^2\ .
  \end{align}
\end{subequations}
This physically plausible result says that the matter energy~$\ED_{(a)}$ of either
particle~$(a=1,2)$ consists of rest mass energy (first terms) plus kinetic energy (second
terms) plus magnetic interaction energy (third terms), while for these single-particle
energies~$\ED_{(a)}$ there appears no electric interaction energy. The emergence of the
\emph{magnetic} kind of interaction energy seems to be somewhat unreasonable; but this is
to be understood as the field theoretic counterpart of the minimal substitution
$(\hat{\vec{p}}\to\hat{\vec{p}}-\frac{e}{c}\vec{A})$ for the conventional energy
$(\hat{H}=\frac{\vec{p\,}^2}{2m})$ of a point particle moving in a magnetic
field~$\vec{H}=\nabla\times\vec{A}$.

The third interesting point with that second proposal~$\tETii$~(\ref{eq:IV.4}) is now that
by use of the explicit form~(\ref{eq:III.19a})-(\ref{eq:III.19b}) of the mass functionals
together with the electric and magnetic Poisson identities,
cf.~(\ref{eq:IV.3a})-(\ref{eq:IV.3b}), this proposal can be rewritten as the sum of the
individual rest mass and kinetic energies plus the gauge field energy of the electric~(e)
and magnetic~(m) type:
\begin{equation}
  \label{eq:IV.7}
  \ETiii = \left(\Z^2_{(1)}\cdot\Mp c^2 + \Z_{(2)}^2\cdot\Me c^2\right) +
  2\left(\Tkin_{(1)}+\Tkin_{(2)}\right) + \left( \heER - \hmER\right)\ .
\end{equation}

Here, the validity of the normalization conditions~(\ref{eq:III.22}) has tacitly been
assumed and therefore they do not appear explicitly in the present result
for~$\ETiii$. Observe however that the physically reasonable form~(\ref{eq:IV.7}) of the
energy functional is a consequence of the fact that its preliminary
form~$\tETii$~(\ref{eq:IV.4}) contains the gauge field energies of electric and magnetic
type with different signs. This important fact is the reason why the \emph{double
  counting} of the electric term~(\ref{eq:IV.3a}) in the sum~$\tMT c^2$~(\ref{eq:III.24}) of
mass eigenvalues~$M_{[a]}c^2$~(\ref{eq:III.19a})-(\ref{eq:III.19b}) becomes
\emph{compensated} (see the discussion of this effect in ref.s~\cite{BeSo,PS}); and then the
electric field energy~$\heER$ appears only once in the third
proposal~$\ETiii$~(\ref{eq:IV.7}). For the magnetic field energy~$\hmER$ there occurs an
analogous effect since its (negative) double-counting in the sum~$\tMT
c^2$~(\ref{eq:IV.4}) is weakened so that the magnetic field energy~$\hmER$ appears now in
the third proposal~(\ref{eq:IV.7}) with the opposite sign relative to its electric
counterpart~$\heER$! This circumstance however does not influence the lowest-order
approximation of the atomic energy levels because these are dominated by the
\emph{electric} interactions. Nevertheless for the higher-order approximations, the
negative sign of the magnetic term will leave its imprint upon the predictions and
therefore must receive confirmation or rejection by the observational data (see below).

The final step for the construction of the wanted energy functional must now be based upon
the somewhat amazing circumstance that the third proposal~$\ETiii$~(\ref{eq:IV.7}) would
numerically produce the same energy upon an exact solution of the RST eigenvalue problem
as does the original functional~$\ET$~(\ref{eq:II.53})-(\ref{eq:II.55b}), too. The reason
is that the transcription of~$\ET$ to~$\ETiii$ relies exclusively upon the use of the
Poisson identities which, however, do automatically hold for any exact solution of the RST
eigenvalue problem! Nevertheless, this third form~$\ETiii$~(\ref{eq:IV.7}) cannot be used
for the deduction of the mass eigenvalue and Poisson equations as the corresponding
variational equations since this functional~(\ref{eq:IV.7}) contains no coupling at all
between the matter fields~$\psi_a(\vec{r})$ and the gauge
fields~$\aAn,\vec{A}_a(\vec{r})$. Indeed, the coupling of matter and gauge fields has been
eliminated on the way from the original~$\ET$~(\ref{eq:II.53}) to the
present~$\ETiii$~(\ref{eq:IV.7}) via the Poisson identities, albeit under simultaneous
preservation of the numerical value of the energy functional.

Therefore it finally becomes necessary to restore that lost coupling of matter and gauge
fields, again under preservation of the numerical value of the energy functional. Naturally
one expects that such a restoration of the desired coupling must be performed with regard again
of the Poisson identities which thus have to take over the role of constraints for the
variational procedure (see also ref.~\cite{BMS}). In this sense, one resorts to the method
of Lagrangean multipliers with respect to both the wave function
normalizations~(\ref{eq:III.22}) and the Poisson
identities~(\ref{eq:III.58})-(\ref{eq:III.59c}); and thus one complements the third
proposal~$\ETiii$~(\ref{eq:IV.7}) to the final result~$\tET$ in the following way:
\begin{equation}
  \label{eq:IV.8}
  \tET = \ETiii + \sum_{a=1}^2\left(\lD_{(a)}\cdot\ND_{(a)}+\lGe_{(a)} \cdot
    \NGe_{(a)} + \lGm_{(a)}\cdot\NGm_{(a)} \right)\ .
\end{equation}
Here it is now a standard exercise to convince oneself of the fact that the variational
equations due to this functional~$\tET$ actually are just the mass eigenvalue
equations~(\ref{eq:III.15}) and~(\ref{eq:III.17}) \emph{together} with the electric and
magnetic Poisson equations~(\ref{eq:III.47a})-(\ref{eq:III.47d}), provided the Lagrangean
matter multipliers~$\lD_{(a)}$ are given in terms of the mass eigenvalues~$M_a$ as shown
by equations~(\ref{eq:III.23a})-(\ref{eq:III.23b}) and furthermore the gauge field
multipliers~$\lG^{\mathrm{(e,m)}}_{(a)}$ are specified as follows~$(a=1,2)$
\begin{subequations}
  \begin{align}
    \label{eq:IV.9a}
    \lGe_{(a)} &= -\frac{\hbar c}{4\pi\as}\\*
    \label{eq:IV.9b}
    \lGm_{(a)} &= \frac{\hbar c}{4\pi\as}\ .
  \end{align}
\end{subequations}
Thus collecting all the partial results, the ultimate form of the wanted energy
functional~$\tET$ is the following:
\begin{equation}
  \label{eq:IV.10}
  \begin{split}
    \tET &= \Z_{(1)}^2\cdot\Mp c^2 + \Z_{(2)}^2\cdot\Me c^2 + 2\left( \Tkin_{(1)} +
      \Tkin_{(2)} \right) +\left( \heER - \hmER \right)\\*
    &+ M_1 c^2\cdot\ND_{(1)} - M_2 c^2\cdot\ND_{(2)} +
    \frac{\hbar c}{4\pi\as}\sum_{a=1}^2\left(\NGm_{(a)}-\NGe_{(a)}\right)\ .
  \end{split}
\end{equation}

The practical usefulness of this ultimate energy functional~$\tET$~(\ref{eq:IV.10}) refers
mainly to those situations where the RST eigenvalue problem cannot be solved exactly so
that one is forced to look for approximate solutions (which will be mostly the case). But
fortunately, a convenient approximation method is now at hand in form of the energy
functional~$\tET$, so that one can test certain trial functions for the Dirac
spinors~$\psi_a(\vec{r})$ and for the gauge potentials~$\aAn,\vec{A}_a(\vec{r})$. These trial
functions will depend upon some parameters~$(b_k)$ so that, after substitution of the
trial functions into the energy functional~$\tET$~(\ref{eq:IV.10}), the latter becomes an
ordinary function of the ansatz parameters~$b_k$:~$\tET=\tET(b_k)$. Finally, looking for
the minimally possible value of that function~$\tET(b_k)$ yields a more or less good
approximation for the wanted energy eigenvalue of the RST eigenvalue problem. For an
example of this type see ref.~\cite{BMS}. However a further improvement of this general
approximation procedure may be achieved by not trying some independent functions for the
gauge potentials~$\aAn$ and~$\vec{A}_a(\vec{r})$ but by trying merely for the (normalized)
wave functions~$\psi_a(\vec{r})$ and then calculating (exactly) the associated gauge
potentials from their Poisson equations~(\ref{eq:III.47a})-(\ref{eq:III.47d}), preferably
in form of the solutions~(\ref{eq:III.48a})-(\ref{eq:III.48d}). Clearly, through such a
procedure the Poisson identities~(\ref{eq:III.58})-(\ref{eq:III.59c}) will be satisfied
\emph{exactly}, though the associated solution of the RST eigenvalue problem is an
\emph{approximation}. But the advantage is here that all the constraints (second line on the
right of equation~(\ref{eq:IV.10})) can be omitted and one can concentrate upon the
physical terms (first line) which effectively is~$\ETiii$~(\ref{eq:IV.7}). Thus it will be
sufficient to look for the minimum of the corresponding function~$\ETiii(b_k)$.
Subsequently we will exemplify this procedure by means of the positronium groundstate.

As a preparation of this groundstate treatment, it is very instructive and convenient to
specify the functional~$\ETiii$ in terms of the wave amplitudes~$\aRpm$ and $\aSpm$
(\ref{eq:III.62a})-(\ref{eq:III.62b}). First, the mass renormalization
factors~$\Z_{(a)}^2$~(\ref{eq:III.11}) are found to be of the following form
\begin{equation}
  \label{eq:IV.11}
  \Z_{(a)}^2 = \frac{1}{2}\int d^2\vec{r}\left(\aRp^2+\aSp^2-\aRm^2-\aSm^2 \right)\ ,
\end{equation}
where the unique wave amplitudes~$\aRpm,\aSpm$ are assumed (for the sake of simplicity) to
depend only upon the radial~$(r)$ and longitudinal~$(\vartheta)$ variables: 
$\aRpm\rt,\aSpm\rt$; and the remaining two-dimensional volume element~$d^2\vec{r}$ is then
given in terms of these variables as
\begin{equation}
  \label{eq:IV.12}
  d^2\vec{r} = rdrd\vartheta\ .
\end{equation}

Next, the kinetic energies~$\Tkin_{(a)}$~(\ref{eq:III.20a})-(\ref{eq:III.20b}) of both
particles\\ $(a=1,2)$ are found to split up into the radial~$(T_r)$ and
longitudinal~$(T_\vartheta)$ part, i.e.
\begin{equation}
  \label{eq:IV.13}
  \Tkin_{(a)} = T_{r(a)} + T_{\vartheta(a)}  
\end{equation}
with the radial part being given by
\begin{equation}
  \label{eq:IV.14}
  \begin{split}
  T_{r(a)} = (-1)^{a-1}\frac{\hbar c}{4}\int d^2\vec{r} \left(
    \aRm\cdot\frac{\partial\aRp}{\partial r} - \frac{\aRp}{r}\cdot
    \frac{\partial(r\,\aRm)}{\partial r}  \right.\\*
    \left.+\;\aSm\cdot\frac{\partial\aSp}{\partial r} -
    \frac{\aSp}{r}\cdot\frac{\partial(r\,\aSm)}{\partial r} \right)\ ,    
  \end{split}
\end{equation}
and analogously the longitudinal part by
\begin{equation}
  \label{eq:IV.15}
  \begin{split}
  T_{\vartheta(a)} = (-1)^{a-1}\frac{\hbar c}{4}\int \frac{d^2\vec{r}}{r}\left( \aRm\cdot
    \frac{\partial\aSp}{\partial\vartheta} - \aSp\cdot\frac{\partial\aRm}{\partial\vartheta}
    \right.\\*
    \left.+\;\aRp\cdot\frac{\partial\aSm}{\partial\vartheta} - \aSm\cdot \frac{\partial\aRp}{\partial\vartheta}
    \right)\ .    
  \end{split}
\end{equation}

Furthermore, the electric and magnetic field energies~$\heER$ and~$\hmER$~(\ref{eq:IV.1b})
read in terms of the gauge potentials~$\aAn$ and~${}^{(a)}A_\phi(\vec{r})$~(\ref{eq:III.65})
\begin{subequations}
  \begin{align}
    \label{eq:IV.16a}
    \heER &= \frac{\hbar c}{2\as}\int d^2\vec{r}\; r\sin\vartheta \left(
      \frac{\partial\,\iAn}{\partial r}\cdot \frac{\partial\,\iiAn}{\partial r} +
      \frac{1}{r^2}\frac{\partial\,\iAn}{\partial \vartheta}\cdot \frac{\partial\,\iiAn}{\partial \vartheta}
    \right)\\*
    \label{eq:IV.16b}
    \begin{split}
    \hmER = \frac{\hbar c}{2\as}\int dr d\vartheta \sin\vartheta \left(
      \frac{\partial}{\partial r}(r {}^{(1)}\!A_\phi)\cdot \frac{\partial}{\partial r}(r {}^{(2)}\!A_\phi)\right.\\*
      \left.+\;\frac{1}{\sin^2\vartheta}\frac{\partial}{\partial\vartheta}(\sin\vartheta\, {}^{(1)}\!A_\phi )\cdot
      \frac{\partial}{\partial\vartheta}(\sin\vartheta\, {}^{(2)}\!A_\phi )
      \right).
    \end{split}
  \end{align}
\end{subequations}
Finally, both kinds of constraints, i.e.\ the normalization conditions~(\ref{eq:II.44})
and the Poisson identities~(\ref{eq:III.58})-(\ref{eq:III.59c}), must also be rewritten
in terms of the wave amplitudes; but it is not necessary to reproduce this here because
for the subsequent treatment of the positronium groundstate we will use trial functions
satisfying a priori all those constraints and therefore we can rely directly upon the
truncated functional~$\ETiii$~(\ref{eq:IV.7})  without loss of accuracy.

Now in order to support the confidence in the established functional
$\tET$~(\ref{eq:IV.10}), one can look for both the mass eigenvalue and Poisson equations
in terms of the wave amplitudes~$\aRpm,\aSpm$ by carrying out the variational
procedure~$(\delta\tET=0)$ just with respect to these wave amplitudes and gauge potentials.
Clearly, one will then actually recover the former mass eigenvalue
equations~(\ref{eq:III.66a})-(\ref{eq:III.66d}) together with the Poisson
equations~(\ref{eq:III.47a})-(\ref{eq:III.47d}). For their magnetic
part~(\ref{eq:III.47c})-(\ref{eq:III.47d}) one may resort for the moment to the special
case of circular flow around the z-axis, cf.~(\ref{eq:III.65}); and in this special case
the magnetic Poisson equations for the azimuthal component~${}^{(a)}A_\phi$ read
\begin{subequations}
  \begin{align}
    \label{eq:IV.17a}
    \Delta {}^{(1)}A_\phi - \frac{{}^{(1)}A_\phi}{r^2\sin^2\vartheta} &=
    -4\pi\as{}^{(1)}k_\phi\\*
    \label{eq:IV.17b}
    \Delta {}^{(2)}A_\phi - \frac{{}^{(2)}A_\phi}{r^2\sin^2\vartheta} &=
    4\pi\as{}^{(2)}k_\phi\ ,
  \end{align}
\end{subequations}
with the circular current components~${}^{(a)}k_\phi$ being given by equation~(\ref{eq:III.64}).

\begin{center}
  \emph{\textbf{B.\ Non-Relativistic Approximation}}
\end{center}

For a first practical test of the present construction of an RST energy
functional~$\tET$~(\ref{eq:IV.10}), it may be sufficient to restrict oneself to the
non-relativistic approximation.Clearly, if such an approximation would fail to meet with
the well-known results of ordinary non-relativistic quantum mechanics, one would not try
to further elaborate the corresponding relativistic situation. Fortunately, the
subsequent demonstration by means of the positronium groundstate points just into the
other direction: The \emph{conventional} groundstate energy can be \emph{exactly} reproduced by an
appropriate trial function for the non-relativistic limit of the RST
functional~$\tET$~(\ref{eq:IV.10}), or~$\ETiii$~(\ref{eq:IV.7}), resp. In order to find
the desired non-relativistic limit of the functional~$\ETiii$, it is merely necessary to
look for the non-relativistic forms of its constituents, i.e.\ rest mass and kinetic
energy and the field energy of the electric~$(\heER)$ and magnetic kind~$(\hmER)$.

Naturally, the non-relativistic situation becomes even further simplified if one restricts
oneself to the spherically symmetric approximation by neglecting the magnetic
interactions. As a matter of course, the electric fields~$\vec{E}_a(\vec{r})$ can easily
be visualized to be spherically symmetric (~$\leadsto$ hedgehog configuration) in
contrast to the magnetic fields which mostly obey a dipole (or higher) symmetry. Therefore
it is favorable to start with the spherically symmetric configurations of the purely
electric type.

Turning here first to the mass eigenvalue equations~(\ref{eq:III.66a})-(\ref{eq:III.66d}),
one usually assumes that the ``negative'' Pauli components~$\aRm,\aSm$ are much smaller
than their ``positive'' counterparts~$\aRp$ and~$\aSp$, so that the non-relativistic form of the
mass eigenvalue equations is obtained by simply eliminating those negative
components~$\aRm,\aSm$~\cite{BeSo2}. The residual eigenvalue equations for the 
positive components of the first particle~$(a=1)$ do appear then in the following form:
\begin{subequations}
  \begin{align}
    \label{eq:IV.18a}
    -\frac{\hbar^2}{2\Mp}\left[ \frac{1}{r}\frac{\partial}{\partial r} \left(r\cdot
        \frac{\partial \iRp}{\partial r}\right) + \frac{1}{r^2}
      \frac{\partial^2 \iRp}{\partial\vartheta^2}\right] + \hbar c\iiAn\cdot\iRp =
      \ES_{(1)}\cdot\iRp \\*
    \label{eq:IV.18b}
    -\frac{\hbar^2}{2\Mp}\left[ \frac{1}{r}\frac{\partial}{\partial r} \left(r\cdot
        \frac{\partial \iSp}{\partial r}\right) + \frac{1}{r^2}
      \frac{\partial^2 \iSp}{\partial\vartheta^2}\right] + \hbar c\iiAn\cdot\iSp = \ES_{(1)}\cdot\iSp\ .
  \end{align}
\end{subequations}
Here, the magnetic interactions are neglected~$({}^{(a)}A_\phi\to 0)$ together with the
relativistic effects because both phenomena are mostly of the same (small) order of
magnitude. Moreover, the (conventional) non-relativistic Schr\"odinger
eigenvalues~$\ES_{(a)}$ are defined as in equations~(\ref{eq:III.30a})-(\ref{eq:III.30b}).
The case of the second particle~$(a=2)$ is not written down because it can easily be
supplied by means of the particle permutation symmetry~$(1\leftrightarrow 2)$, see
ref.~\cite{BeSo2}. However the important points with the non-relativistic eigenvalue
equations~(\ref{eq:IV.18a})-(\ref{eq:IV.18b}) refer now to the facts that (i)~they are not
of the usual Schr\"odinger form (\ref{eq:III.32a})-(\ref{eq:III.32b}) and (ii)~the
spin-up~$(\sim\tilde{R}_+)$ and spin-down~$(\sim\tilde{S}_+)$ configurations are
decoupled. The latter circumstance admits us to conceive either of the two single-particle
spins to point definitely into the positive or negative z-direction and their combination
to the para- and ortho-states of the two-particle system will then intuitively be evident.

Naturally, one expects that these non-relativistic eigenvalue equations, such
as~(\ref{eq:IV.18a})-(\ref{eq:IV.18b}), should emerge as the variational equations due
to the non-relativistic approximation~($\tEnT$, say) of the original RST energy
functional~$\tET$~(\ref{eq:IV.10}). Indeed, one is easily convinced that this supposition
is true; namely the elimination of the negative Pauli components~$\aRm,\aSm$ from the
relativistic kinetic energies~$\Tkin_{(a)}$~(\ref{eq:IV.13})-(\ref{eq:IV.15})
yields~\cite{BMS}
\begin{subequations}
  \begin{align}
    \label{eq:IV.19a}
    (\Z_{(1)}^2-1)\Mp c^2 + 2\Tkin_{(1)} &\Rightarrow \Ekin_{(1)} + \EW_{(1)}\\*
    \label{eq:IV.19b}
    (\Z_{(2)}^2-1)\Me c^2 + 2\Tkin_{(2)} &\Rightarrow \Ekin_{(2)} + \EW_{(2)}
  \end{align}
\end{subequations}
with the non-relativistic kinetic energies~$\Ekin_{(a)}$ being given by
\begin{subequations}
  \begin{align}
    \label{eq:IV.20a}
    \Ekin_{(1)} &= \frac{\hbar^2}{4\Mp}\int d^2\vec{r}\left[
      \left(\frac{\partial \iRp}{\partial r}\right)^2 + \frac{1}{r^2}
      \left(\frac{\partial \iRp}{\partial \vartheta}\right)^2\right]\\*
    \label{eq:IV.20b}
    \Ekin_{(2)} &= \frac{\hbar^2}{4\Me}\int d^2\vec{r}\left[
      \left(\frac{\partial \iiSp}{\partial r}\right)^2 + \frac{1}{r^2}
      \left(\frac{\partial \iiSp}{\partial \vartheta}\right)^2\right]\ ,
  \end{align}
\end{subequations}
and the ``winding energies''~$\EW_{(a)}$ being given by
\begin{subequations}
  \begin{align}
    \label{eq:IV.21a}
    \EW_{(1)} &= \frac{\hbar^2}{4\Mp}\int\frac{d^2\vec{r}}{r} \left[
      \frac{\partial \iRp}{\partial r}\cdot\frac{\partial \iSp}{\partial \vartheta} -
      \frac{\partial \iSp}{\partial r}\cdot\frac{\partial \iRp}{\partial \vartheta}\right]\\*
    \label{eq:IV.21b}
    \EW_{(2)} &= \frac{\hbar^2}{4\Me}\int\frac{d^2\vec{r}}{r} \left[
      \frac{\partial \iiRp}{\partial r}\cdot\frac{\partial \iiSp}{\partial \vartheta} -
      \frac{\partial \iiSp}{\partial r}\cdot\frac{\partial \iiRp}{\partial
        \vartheta}\right]\ .
  \end{align}
\end{subequations}
Observe here that, for the kinetic energies~$\Ekin_{(a)}$, we made use of the
non-relativistic decoupling of the spin-up and spin-down components and thus adopted the
first spin~(\ref{eq:IV.18a}) pointing in the positive z-direction~$(\leadsto\iRp)$ and the
second spin~(\ref{eq:IV.18b}) in the negative z-direction~$(\leadsto \iiSp)$. Clearly, the
other combinations of the spin directions~$\{\iRp,\iiRp\}$, $\{\iSp,\iiRp\}$,
$\{\iSp,\iiSp \}$ are equally well possible, see below for the para- and
ortho-configurations. Fortunately, through this choice of definite spin directions for any
particle, the winding energies~(\ref{eq:IV.21a})-(\ref{eq:IV.21b}) become zero so that one
can restrict oneself to the kinetic energies~(\ref{eq:IV.20a})-(\ref{eq:IV.20b}) alone. Of
course the rest mass energies need not be taken into account for a non-relativistic
treatment and therefore have been omitted, cf~(\ref{eq:IV.19a})-(\ref{eq:IV.19b}).

Next, the non-relativistic form of the electric and magnetic gauge field energies~$\heER$
and~$\hmER$ remains the same as in the relativistic case, i.e.\ in terms of the static
gauge potentials~$\aAn$ and~$\vec{A}_a(\vec{r})$
(cf.~(\ref{eq:III.52a})-(\ref{eq:III.52b}):
\begin{subequations}
  \begin{align}
    \label{eq:IV.22a}
    \heER &= \frac{\hbar c}{4\pi\as}\int d^3\vec{r}\;\left(\vec{\nabla}\iAn\right)\cdot
    \left(\vec{\nabla}\iiAn\right)\\*    
    \label{eq:IV.22b}
    \hmER &= \frac{\hbar c}{4\pi\as}\int d^3\vec{r}\;\left(\vec{\nabla}\times\vec{A}_1\right)\cdot
    \left(\vec{\nabla}\times\vec{A}_2\right)\ .
  \end{align}
\end{subequations}
But for the explicit calculation of the non-relativistic potentials~$\aAn$
and~${}^{(a)}A_\phi(\vec{r})$ from the Poisson equations
(\ref{eq:III.47a})-(\ref{eq:III.47b}) and (\ref{eq:III.47c})-(\ref{eq:III.47d}) one will
use the corresponding non-relativistic approximations for the charge and current
densities, cf.~(\ref{eq:III.63a}) and~(\ref{eq:III.64})
\begin{subequations}
  \begin{align}
    \label{eq:IV.23a}
    \ikn &\doteqdot \frac{\itkn\rt}{4\pi r\sin\vartheta} \Rightarrow
    \frac{\iRp^2\rt}{4\pi
      r\sin\vartheta}\\*[5mm]
    \label{eq:IV.23b}
    \iikn &\doteqdot \frac{\iitkn\rt}{4\pi r\sin\vartheta} \Rightarrow
    \frac{\iiSp^2\rt}{4\pi r\sin\vartheta}\\*[5mm]
    \label{eq:IV.23c}
    \ikp &\doteqdot \frac{\itkp\rt}{2\pi r\sin\vartheta} \Rightarrow
    \frac{\iRp\cdot\iRm}{2\pi r}\\*[5mm]
    \label{eq:IV.23d}
    \iikp &\doteqdot \frac{\iitkp\rt}{2\pi r\sin\vartheta} \Rightarrow
    -\frac{\iiSp\cdot\iiSm}{2\pi r}\ .
  \end{align}
\end{subequations}
Observe here again that, in the contrast to the charge densities~$\akn$, the current
densities~$\akp$ are built up by both the positive~$(\aRp,\aSp)$ and
negative~$(\aRm,\aSm)$ wave amplitudes while, properly speaking, the negative
amplitudes~$\aRm,\aSm$ should be neglected against their positive
counterparts~$\iRp,\iSp$ for the non-relativistic limit. This demonstrates that it appears
somewhat inconsequent to retain the magnetic (i.e.\ spin-spin) interactions for the
non-relativistic approximation because their order of magnitude may be the same as that of
the other dominant relativistic effects. Nevertheless we will not drop the magnetic
effects for our non-relativistic approach because one can still deal with the effect of
ortho-para splitting of the energy levels from a more qualitative viewpoint.

In this sense one has to renounce on the inclusion of the magnetic effects for the purpose
of deducing the non-relativistic eigenvalue equations (\ref{eq:IV.18a})-(\ref{eq:IV.18b})
from the desired non-relativistic version~$\tEnT$ of the original functional~$\tET$
(\ref{eq:IV.10}). Consequently one drops also the magnetic
constraints~$\NGm_{(a)}$~(\ref{eq:III.59b})-(\ref{eq:III.59c}) from the latter functional
and retains only the electric constraints
$\NGe_{(a)}$~(\ref{eq:III.58})-(\ref{eq:III.59a}) which do then appear in the following
form:
\begin{subequations}
  \begin{align}
    \label{eq:IV.24a}
    \NGe_{(1)} &\Rightarrow \NGn_{(1)} = \int d^3\vec{r}\left[
      \left(\vec{\nabla}\iAn\right)\cdot\left(\vec{\nabla}\iiAn\right) +
      \as\frac{\iAn\cdot\iiSp^2}{r\sin\vartheta} \right]\\*
    \label{eq:IV.24b}
    \NGe_{(2)} &\Rightarrow \NGn_{(2)} = \int d^3\vec{r}\left[
      \left(\vec{\nabla}\iAn\right)\cdot\left(\vec{\nabla}\iiAn\right) -
      \as\frac{\iiAn\cdot\iRp^2}{r\sin\vartheta} \right]\ ,
  \end{align}
\end{subequations}
where the non-relativistic approximations~(\ref{eq:IV.23a})-(\ref{eq:IV.23b}) of the
charge densities~$\akn$ have already been respected. Clearly, the latter approximations
for the charge densities must also be used for the constraints of wave function
normalization~(\ref{eq:III.22}) which then appear in their following non-relativistic
forms~$\NDn_{(a)}$:
\begin{subequations}
  \begin{align}
    \label{eq:IV.25a}
    \ND_{(1)} &\Rightarrow \NDn_{(1)} \doteqdot \frac{1}{2}\int d^2\vec{r}\;\;\iRp^2\rt -1 = 0\\*
    \label{eq:IV.25b}
    \ND_{(2)} &\Rightarrow \NDn_{(2)} \doteqdot \frac{1}{2}\int d^2\vec{r}\;\;\iiSp^2\rt
    -1 = 0\ .
  \end{align}
\end{subequations}

Finally, collecting all the non-relativistic approximations and applying again the method
of Lagrangean multipliers lets appear the wanted non-relativistic approximation~$\tEnT$ of
the original functional~$\tET$~(\ref{eq:IV.10}) in the following form:
\begin{equation}
  \label{eq:IV.26}
  \begin{split}
    \tEnT &= \Ekin_{(1)}  + \Ekin_{(2)} + \heER \\*
    &+ \sum_{a=1}^2\lS_{(a)}\cdot\NDn_{(a)}-\frac{\hbar c}{4\pi\as}\sum_{a=1}^2\NGn_{(a)}\ .
  \end{split}
\end{equation}
Here it is again a nice consistency check to convince oneself of the fact that the usual
variational procedure~$(\delta\tEnT=0)$ actually does reproduce the claimed
non-relativistic forms (\ref{eq:IV.18a})-(\ref{eq:IV.18b}) and
(\ref{eq:III.47a})-(\ref{eq:III.47b}) of the mass eigenvalue and Poisson equations. The
non-relativistic multipliers~$\lS_{(a)}$ turn out as the conventional Schr\"odinger
energie eigenvalues
\begin{equation}
  \label{eq:IV.27}
  \lS_{(a)} = -\ES_{(a)}\ ,
\end{equation}
which compares to the analogous result~(\ref{eq:III.40}) of the Ritz-Hardy-Schr\"odinger
approach. Clearly according to our present choice of the negative z-direction for the
second particle spin, the second eigenvalue equation~(\ref{eq:IV.18b}) for the first
particle must be replaced for the present situation by
\begin{equation}
  \label{eq:IV.28}
  -\frac{\hbar^2}{2Me}\left[\frac{1}{r}\frac{\partial}{\partial r} \left( r\cdot
      \frac{\partial \iiSp}{\partial r} \right) + \frac{1}{r^2}\frac{\partial^2
      \iiSp}{\partial\vartheta^2}\right] - \hbar c\; {}^{(1)}\!A_0\cdot\iiSp = -\lS_{(2)}\cdot\iiSp
\end{equation}
for the second particle. Furthermore, the non-relativistic approximations of the electric
Poisson equations do now appear as the following variational equations:
\begin{subequations}
  \begin{align}
    \label{eq:IV.29a}
    \Delta\,  {}^{(1)}\!A_0 &= -\as\frac{\iRp^2}{r\sin\vartheta}\\*
    \label{eq:IV.29b}
    \Delta\,  {}^{(2)}\!A_0 &= \as\frac{\iiSp^2}{r\sin\vartheta}\ .
  \end{align}
\end{subequations}

The important point here is that, despite the many similarities between the
non-relativistic limit of RST and the conventional Ritz-Hartree-Schr\"odinger approach,
there are also characteristic differences of both approaches which are in favour of
RST. This will readily be demonstrated by considering a numerical example. The main
difference refers to the gauge potentials, e.g.\ those of the electric
type~(\ref{eq:III.48a})-(\ref{eq:III.48b}), which by means of the non-relativistic
approximations~(\ref{eq:IV.23a})-(\ref{eq:IV.23b}) for the charge densities~$\akn$ appear
as 
\begin{subequations}
  \begin{align}
    \label{eq:IV.30a}
    \iAn &=
    \frac{\as}{4\pi}\int\frac{d^3\vec{r}\,'}{r'\sin\vartheta'} \cdot
    \frac{\iRp^2(r',\vartheta')}{||\vec{r}-\vec{r}\,' ||}\\*
    \label{eq:IV.30b}
    \iiAn &=
    -\frac{\as}{4\pi}\int\frac{d^3\vec{r}\,'}{r'\sin\vartheta'}\cdot\frac{\iiSp^2(r',\vartheta')}
    {||\vec{r}-\vec{r}\,' ||}\ .
  \end{align}
\end{subequations}
Obviously, these gauge potentials due to the exotic states must be more singular as the
Hartree potentials~(\ref{eq:III.50a})-(\ref{eq:III.50b}) which are due to the \emph{non-singular}
Hartree wave functions~$\varphi_a(\vec{r})$~(\ref{eq:III.31}), see the example (\ref{eq:III.73}).

\begin{center}
  \emph{\textbf{C.\ Magnetic Interactions}}
\end{center}

In the contrast to the electric fields~$\vec{E}_a(\vec{r})$~(\ref{eq:III.52a}), the
magnetic fields $\vec{H}_a(\vec{r})$ (\ref{eq:III.52b}) cannot obey the SO(3) symmetry
because they have dipole character rather than monopole character like their electric
counterparts. Therefore it will become necessary to apply more complicated approximation
techniques; but fortunately it is not necessary to explicitly solve the magnetic Poisson
equations~(\ref{eq:III.47c})-(\ref{eq:III.47d}) for the three-vector potentials
$\vec{A}_a(\vec{r})$, e.g.\ in form of the special solutions
(\ref{eq:III.48c})-(\ref{eq:III.48d}). Rather it is sufficient to determine the magnetic
fields~$\vec{H}_a(\vec{r})\ (a=1,2)$ directly from the Abelian Maxwell equations
\begin{equation}
  \label{eq:IV.31}
  \vec{\nabla}\times\vec{H}_a = 4\pi\as\vec{j}_a\ ,
\end{equation}
which is the three-vector form of the relativistic versions
(\ref{eq:II.27a})-(\ref{eq:II.27b}). Nevertheless, one has to insist on the existence of the
corresponding vector potentials~$\vec{A}_a(\vec{r})$ (\ref{eq:III.52b}), namely in order
that the magnetic Poisson identities (\ref{eq:III.59b})-(\ref{eq:III.59c}) can be
satisfied and thus the corresponding magnetic constraints in the energy functional~$\tET$
(\ref{eq:IV.10}) can be dropped. Indeed, in the latter case one can restrict oneself to the
physical terms of the energy~$\tET$ (i.e.\ the first line on the right-hand side of
(\ref{eq:IV.10})), where the magnetic interaction energy is then simply given by~$\hmER$
(\ref{eq:IV.1b}) in terms of the magnetic fields~$\vec{H}_a$ themselves. But clearly if
there is no difficulty with the determination of the vector potentials~$A_a(\vec{r})$
(\ref{eq:III.52b}) directly from their Poisson equations, one may calculate the magnetic
interaction energy~$\hmER$ also in terms of these vector potentials~$\vec{A}_a(\vec{r})$
as shown by equation (\ref{eq:IV.22b}).

Following here the first path (i.e.\ determination of the magnetic
fields~$\vec{H}_a(\vec{r})$ directly from the Maxwell equations (\ref{eq:IV.31}) with
omission of the vector potentials~$\vec{A}_a(\vec{r})$), one additionally has to impose
the conditions of vanishing sources
\begin{equation}
  \label{eq:IV.32}
  \vec{\nabla}\sdot\vec{H}_a(\vec{r})=0\ ,
\end{equation}
in order to ensure the existence of the vector potentials~$\vec{A}_a(\vec{r})$. Since the
Maxwell currents~$\vec{j}_a(\vec{r})$ are connected to the Dirac
currents~$\vec{k}_a(\vec{r})$ by equations (\ref{eq:II.43a})-(\ref{eq:II.43b}), the
Abelian Maxwell equations (\ref{eq:IV.31}) read in the component form of the spherical
polar coordinates~$(a=1,2)$
\begin{equation}
  \label{eq:IV.33}
  \frac{1}{r}\left[\frac{\partial}{\partial r}\left(r\cdot {}^{(a)}H_\vartheta\right)
    -\frac{\partial {}^{(a)}H_r}{\partial\vartheta} \right] =
  4\pi\as {}^{(a)}k_\phi\ ,
\end{equation}
where the azimuthal components~${}^{(a)}k_\phi$ of the Dirac currents~$\vec{k}_a(\vec{r})$
are specified by equations (\ref{eq:IV.23c})-(\ref{eq:IV.23d}). However for the present
non-relativistic limit, the ``negative'' Pauli wave amplitudes~$\aRm,\aSm$ must be
approximately traced back to their ``positive'' counterparts~$\aRp,\aSp$
through~\cite{BMS,BeSo2}
\begin{subequations}
  \begin{align}
    \label{eq:IV.34a}
    \iRp &\simeq \frac{\hbar}{2\Mp c} \left(\frac{\partial\iRp}{\partial r} +
      \frac{1}{r}\frac{\partial\iSp}{\partial\vartheta} \right)\Rightarrow
    \frac{\hbar}{2\Mp c}\frac{\partial\iRp}{\partial r}\\*
    \label{eq:IV.34b}
    \iSp &\simeq \frac{\hbar}{2\Mp c} \left(\frac{\partial\iSp}{\partial r} -
      \frac{1}{r}\frac{\partial\iRp}{\partial\vartheta} \right)\Rightarrow
    \frac{\hbar}{2\Mp c}\frac{\partial\iSp}{\partial r}
  \end{align}
\end{subequations}
and analogously for the second particle~$(1\to 2;\Mp\to\Me)$. This approximative procedure
recasts the non-relativistic current components~${}^{(a)}k_\phi$
(\ref{eq:IV.23c})-(\ref{eq:IV.23d}) to the following form:
\begin{subequations}
  \begin{align}
    \label{eq:IV.35a}
    {}^{(1)}k_\phi &\simeq \frac{\hbar}{2\Mp}\cdot\frac{\iRp\cdot
      \frac{\ds \partial}{\ds \partial r}\iRp}{2\pi r}\\*
    \label{eq:IV.35b}
    {}^{(2)}k_\phi &\simeq -\frac{\hbar}{2\Me}\cdot\frac{\iiSp\cdot
      \frac{\ds\partial}{\ds\partial r}\iiSp}{2\pi r}\ .
  \end{align}
\end{subequations}
Evidently, the currents~$\vec{k}_a(\vec{r})$ become singular at the origin~$(r\to 0)$
unless the wave amplitudes~$\iRp,\iSp$ or their derivatives do vanish for~$r\to 0$.

Of course, these singular currents will imply a corresponding singular behavior of the
magnetic components~$ {}^{(a)}H_r$ and~$ {}^{(a)}H_\vartheta$ as solutions of the Maxwell
equations (\ref{eq:IV.33}). In order to elaborate this singular behavior in some more
detail, one splits off the short-range magnetic field~$\vec{H}_a'(\vec{r})$ from its
far-range counterpart which can always be taken as a gradient field, i.e.\ we put
\begin{equation}
  \label{eq:IV.36}
  \vec{H}_a(\vec{r}) = \vec{H}'_a(r) + \vec{\nabla}\eta_{(a)}\ .
\end{equation}
The inclusion of such a gradient field is necessary in order to satisfy the divergence
relation (\ref{eq:IV.32}) which yields for the \emph{magnetic potentials}~$\eta_{(a)}$
\begin{equation}
  \label{eq:IV.37}
  \Delta\eta_{(a)} = - \left(\vec{\nabla}\cdot\vec{H}'_a\right)\ .
\end{equation}
Thus the magnetic Maxwell equations (\ref{eq:IV.31}) do fix only the curl of the
short-range fields~$\vec{H}'_a(\vec{r})$ and leave the determination of the magnetic
potentials~$\eta(a)$ to the Poisson equation (\ref{eq:IV.37}).

Observe also that the superposition of a magnetic potential~$\eta_{(a)}$ to the original
magnetic field~$\vec{H}'_a(r)$ (\ref{eq:IV.36}), so that the resulting
field~$\vec{H}_a(\vec{r})$ gets vanishing divergence (\ref{eq:IV.32}), does not only
ensure the existence of a vector potential~$\vec{A}_a(\vec{r})$
for~$\vec{H}_a(=\vec{\nabla}\times\vec{A}_a)$  but additionally implies a further effect
which refers to the magnetic field energy~$\hmER$ (\ref{eq:IV.1b}). Indeed the latter
object reads by use of the superposition (\ref{eq:IV.36})
\begin{equation}
  \label{eq:IV.38}
  \begin{split}
    \hmER &= \frac{\hbar c}{4\pi\as}\int d^3\vec{r}\;\vec{H}_1(\vec{r})\sdot
    \vec{H}_2(\vec{r})\\*
    &= \frac{\hbar c}{4\pi\as}\int d^3\vec{r}\; \left( \vec{H}_1'\sdot\vec{H}_2'
      + \vec{H}_1'\sdot\vec{\nabla}\eta_{(2)} +  \vec{H}_2'\sdot\vec{\nabla}\eta_{(1)}
      + \vec{\nabla}\eta_{(1)}\sdot\vec{\nabla}\eta_{(2)}\right)\ .
  \end{split}
\end{equation}
Here it is easy to see that for given fields~$\vec{H}_a'(r)$ this functional of the
magnetic potentials~$\eta_{(a)}$ is stationary just upon the solutions of the Poisson
equations (\ref{eq:IV.37})! Thus the introduction of the magnetic potentials~$\eta_{(a)}$
does not only guarantee the existence of the vector potentials~$\vec{A}_a(\vec{r})$ but it
additionally lets the magnetic energy functional~$\hmER$ appear stationary (independent of
the stationarity of the total functional~$\tET$). 

The ansatz for the components of the short-range fields~$\vec{H}'_a$ is now
\mbox{$(a=1,2)$}:
\begin{subequations}
  \begin{align}
    \label{eq:IV.39a}
    {}^{(a)}\!H'_r(r,\vartheta) &= {}^{(a)}\!H'_\varphi(r,\vartheta)\equiv 0\\*
    \label{eq:IV.39b}
    {}^{(a)}\!H'_\vartheta(r,\vartheta) &= h_a(r)\ ,
  \end{align}
\end{subequations}
i.e.\ the integral lines of the short-range fields~$\vec{H}_a'=h_a(r)\vec{e}_\vartheta$
are circles \mbox{$r=const.$} in the two-planes~$\phi=const.$ with the center located at
the origin~$r=0$. Clearly, such a field must necessarily be singular along the z-axis but
this singularity does not contribute to the magnetic field energy~$\hmER$.

The magnetic ansatz functions~$h_a(r)$ are linked to the azimuthal components
${}^{(a)}k_\phi(r)$ (\ref{eq:IV.35a})-(\ref{eq:IV.35b}) via the Maxwell equations
(\ref{eq:IV.33}), which do apply also to the short-range
components~${}^{(a)}H_r',{}^{(a)}H_\vartheta'$, yielding
\begin{subequations}
  \begin{align}
    \label{eq:IV.40a}
    \frac{1}{r}\frac{d}{dr}\left(rh_1\right) = 4\pi\as {}^{(1)}k_\phi\\*
    \label{eq:IV.40b}
    \frac{1}{r}\frac{d}{dr}\left(rh_2\right) = -4\pi\as {}^{(2)}k_\phi\ .
  \end{align}
\end{subequations}
But since the current components~$ {}^{(a)}k_\phi$ are just of that specific form
(\ref{eq:IV.35a})-(\ref{eq:IV.35b}), the solutions of the present differential equations
(\ref{eq:IV.40a})-(\ref{eq:IV.40b}) are easily found as
\begin{subequations}
  \begin{align}
    \label{eq:IV.41a}
    h_1(r) &= \frac{\as\hbar}{2\Mp c}\cdot\frac{\iRp(r)^2}{r}\\*
    \label{eq:IV.41b}
    h_2(r) &= \frac{\as\hbar}{2\Me c}\cdot\frac{\iiSp(r)^2}{r}\ ,
  \end{align}
\end{subequations}
i.e.\ the short-range fields~$\vec{H}_a'(r)=\left\{{}^{(a)}H_r',{}^{(a)}H_\vartheta'
\right\}$  (\ref{eq:IV.39a})-(\ref{eq:IV.39b}) can be directly traced back to the
non-relativistic wave amplitudes~$\iRp,\iiSp$.

Unfortunately, the determination of the long-range magnetic potentials~$\eta_{(a)}$ from
their Poisson equations (\ref{eq:IV.37}) is technically somewhat more complicated. It is
true, the source of the short-range fields~$\vec{H}_a'(\vec{r})$ is relatively simple
\begin{equation}
  \label{eq:IV.42}
  \vec{\nabla}\sdot\vec{H}'_a(r) = \frac{\cot\vartheta}{r}\cdot h_a(r)  \ .
\end{equation}
Thus the corresponding standard solutions for the magnetic potentials~$\eta_{(a)}$ are
given by
\begin{equation}
  \label{eq:IV.43}
  \eta_{(a)} \doteqdot {}^{(a)}\eta(r,\vartheta) = \frac{1}{4\pi}
  \int\frac{d^3\vec{r}\,'}{r'}\frac{\cot\vartheta'\cdot h_a(r')}{||\vec{r}-\vec{r}\,' ||}\ ,
\end{equation}
where the radial functions~$h_a(r)$ are specified by equations
(\ref{eq:IV.41a})-(\ref{eq:IV.41b}). However it seems here that the integral cannot be
calculated in terms of analytic functions, not even for the simple exponential trial
form (\ref{eq:III.68}) for the wave amplitudes~$\iRp^2,\iiSp^2$. Therefore one will be forced
to apply more or less effective approximation methods. For the present purpose one
expands the denominator in the integral (\ref{eq:IV.43}) as follows
\begin{equation}
  \label{eq:IV.44}
  \frac{1}{||\vec{r}-\vec{r}\,' ||} = \frac{1}{\sqrt{r^2+r'^2}}\cdot \left[
    1+\frac{\vec{r}\cdot\vec{r}\,'}{r^2+r'^2} + \frac{3}{2}
    \left(\frac{\vec{r}\cdot\vec{r}\,'}{r^2+r'^2} \right)^2 + \ldots
  \right]\ .
\end{equation}
Here the first (i.e.\ monopole) term does not contribute to the magnetic
potential~$\eta_{(a)}$ (\ref{eq:IV.43}) so that we may be satisfied in the lowest order
with the dipole approximation (second term). Thus the magnetic potential~$\eta_{(a)}$
becomes then in this dipole approximation
\begin{equation}
  \label{eq:IV.45}
  \eta_{(a)}\Rightarrow {}^{(D)}\eta_a(r,\vartheta) = \frac{\pi}{4} r\cos\vartheta
  \int_0^\infty dr'\;r'^2 \frac{h_a(r')}{\sqrt{r^2+r'^2}^3}\ .
\end{equation}

The dipole character of this result becomes evident from its asymptotic
behavior~$(r\to\infty)$:
\begin{equation}
  \label{eq:IV.46}
   {}^{(D)}\eta_a(r\to\infty,\vartheta) =
   \frac{\pi}{4}\cdot\frac{\cos\vartheta}{r^2}\int_0^\infty dr'\;r'^2 h_a(r')\ .
\end{equation}
If the preceding results (\ref{eq:IV.41a})-(\ref{eq:IV.41b}) are used here, with
observation of the non-relativistic normalization conditions (\ref{eq:III.69}) reading
explicitly, e.g., for the second particle
\begin{equation}
  \label{eq:IV.47}
  \int_0^\infty dr\;r\;\iiSp^2(r) = \frac{2}{\pi}\ ,
\end{equation}
then the second magnetic potential (\ref{eq:IV.43}) appears in the asymptotic
region~$(r\to\infty)$ as
\begin{equation}
  \label{eq:IV.48}
   {}^{(D)}\eta_2(r\to\infty,\vartheta) = \frac{e}{\hbar c} \cdot \frac{\mu_{\mathrm{B}}}{2}
   \cdot\frac{\cos\vartheta}{r^2}\ .
\end{equation}

Apart from the dimensional factor~$(\frac{e}{\hbar c})$, which is due to our use of
geometric units for the potentials and field strengths (see equation (\ref{eq:II.11})),
the present result (\ref{eq:IV.48}) for the asymptotic magnetic potential is the usual one
for a magnetic dipole which however carries only \emph{half} of a Bohr
magneton~$\mu_{\mathrm{B}}(\doteqdot\frac{e\hbar}{2\Me c})$. Clearly, this is a further
unconventional feature of the exotic wave functions~$\psi_a(\vec{r})$; namely besides
their singular character for~$r\to 0$ (\ref{eq:III.62a})-(\ref{eq:III.62b}), their
doubled-valuedness (\ref{eq:III.60}), and their integer spin eigenvalue (\ref{eq:III.61}).

It must be stressed, however, that these exotic states do not induce any 
pathological feature into the theory, neither with respect to the electric
field nor for its magnetic counterpart. As a brief demonstration one may
inspect the magnetic fields in the vicinity of the origin~$(r\to 0)$. First, the
scalar magnetic potential~$\eta_{(2)}$ is rewritten as
\begin{equation}
  \label{eq:IV.49}
   {}^{(D)}\eta_2(r,\vartheta) = \frac{e\mu_{\mathrm{B}}}{2\hbar
     c}\cdot\frac{\cos\vartheta}{r^2} g_2(r)\ ,
\end{equation}
with the dipole screening factor~$g_2(r)$ being given by
\begin{equation}
  \label{eq:IV.50}
  g_2(r) = \pi\frac{\Me c}{\as\hbar}\, r^3\int_0^\infty dr'\;
  \frac{r'^2\cdot h_2(r')}{\sqrt{{r^2+r'^2}}^3}\stackrel{r\to\infty}{\to}1\ ,
\end{equation}
so that the asymptotic dipole behavior (\ref{eq:IV.48}) is immediately manifest. (For the
first particle,~$a=1$, the same arguments do hold with merely the electron mass~$\Me$
being replaced by the mass~$\Mp$ of the positive particle). But the crucial point with the
scalar magnetic potentials~${}^{(D)}\eta_2(r,\vartheta)$ (\ref{eq:IV.45}) is now that they do
not diverge at the origin. One is easily convinced of this assertion by tentatively
substituting for the magnetic ansatz function~$h_2(r)$ its non-relativistic form
(\ref{eq:IV.41b}) with the wave amplitude~$\iiSp$ being deduced from the charge
density~${}^{(b)}\tilde{k}_0(r)$ (\ref{eq:III.68}) as
\begin{equation}
  \label{eq:IV.51}
  \iiSp(r) = \sqrt{\frac{8}{\pi r_*^2}}\cdot\exp\left(-\frac{r}{r_*} \right)\ .
\end{equation}
This then yields for the dipole screening factor~$g_2(r)$ (\ref{eq:IV.50})
\begin{equation}
  \label{eq:IV.52}
  g_2(r) = \left(\frac{2}{r_*}\right)^2 r^3\cdot\int_0^\infty
  dr'\;\frac{r'\exp\left(-\frac{\ds 2r'}{\ds r_*}\right)} {\sqrt{{r^2+r'^2}}}\ ,
\end{equation}
which by substitution of the integration variable~$r'$
\begin{equation}
  \label{eq:IV.53}
  \rho\doteqdot\frac{r'}{r}
\end{equation}
adopts the following form
\begin{equation}
  \label{eq:IV.54}
  g_2(r) = \left(\frac{2}{r_*}\right)^2 r^2\int_0^\infty d\rho\;
  \frac{\rho\cdot\exp\left(-\frac{\ds 2r}{\ds r_*}\cdot\rho \right)}
  {\sqrt{1+\rho^2}^3}\ .
\end{equation}
However in this form, it is easy to see that in the vicinity of the origin~$(r\to 0)$ the
screening factor~$g_2(r)$ looks as follows
\begin{equation}
  \label{eq:IV.55}
  g_2(r)\to  \left(\frac{2}{r_*}\right)^2\cdot r^2
\end{equation}
and thus yields a finite value of the magnetic dipole potential~${}^{(D)}\eta_2(r,\vartheta)$
(\ref{eq:IV.49}) around the origin. This compares to the analogous behavior of the
electric potential~${}^{(a)}\!A_0(r)$, see the discussion below equation (\ref{eq:III.73}).

But once it is guaranteed that the vector potentials~$\vec{A}_a(\vec{r})$ do really exist,
one can use this fact in order to recast the magnetic interaction energy~$\hmER$
(\ref{eq:IV.1b}) in a new form which exclusively is based upon the Dirac
currents~$\vec{k}_a(\vec{r})$:
\begin{equation}
  \label{eq:IV.56}
  \hmER = - e^2 \iint d^3\vec{r}_1 d^3\vec{r}_2\;\frac{\vec{k}_1(\vec{r}_1)\sdot
    \vec{k}_2(\vec{r}_2)}{||\vec{r}_1-\vec{r}_2 ||}\ .
\end{equation}
Indeed in order to arrive at this result, one merely has to substitute the vector
potentials~$\vec{A}_a(\vec{r})$ in the magnetic mass equivalents~$\Mmi c^2$ or~$\Mmii c^2$
of the Poisson identities (\ref{eq:IV.3b}) by the formal solution
(\ref{eq:III.48c})-(\ref{eq:III.48d}). Furthermore, both Dirac
currents~$\vec{k}_a(\vec{r})$ are of the azimuthal form (\ref{eq:III.63b})
with~${}^{(a)}k_\phi$ being specified by equations (\ref{eq:IV.23c})-(\ref{eq:IV.23d}),
which in their non-relativistic form appear as shown by equations
(\ref{eq:IV.35a})-(\ref{eq:IV.35b}). Thus the magnetic energy~$\hmER$ does finally emerge
in the following form for identical rest masses~$(\Me=\Mp\doteqdot M)$
\begin{equation}
  \label{eq:IV.57}
  \hmER = \left(\frac{e\hbar}{8\pi Mc}\right)^2\iint \frac{d^3\vec{r}_1}{r_1}\cdot
  \frac{d^3\vec{r}_2}{r_2}\;
  \frac{\ds\frac{d}{dr_1}\left(\iRp(r_1)\right)^2\cdot \frac{d}{dr_2}
    \left(\iiSp(r_2)\right)^2}{||\vec{r}_1-\vec{r}_2||}\ .
\end{equation}
Recalling here the fact that for the groundstate both particles must be in the same
quantum state (apart from the spin direction), one puts
\begin{subequations}
  \begin{align}
    \label{eq:IV.58a}
    \iRp(r) = \iiSp(r) \Rightarrow \tilde{R}(r) &=
    \sqrt{{}^{(b)}\tilde{k}_0(r)}=\sqrt{ \frac{8}{\pi
        r_*^2}}\cdot\exp\left[-\frac{r}{r_*}\right] \ ,\\*
    \label{eq:IV.58b}
    \iSp &= \iiRp \equiv 0
  \end{align}
\end{subequations}
and thus the magnetic energy (\ref{eq:IV.57}) becomes
\begin{equation}
  \label{eq:IV.59}
  \begin{split}
   \hmER &= \left(\frac{2e\hbar}{\pi^2 Mcr^3_*}\right)^2 \iint  \frac{d^3\vec{r}_1}{r_1}\cdot
   \frac{d^3\vec{r}_2}{r_2}\;\frac{ \exp\left[-\frac{2}{r_*}\left(r_1+r_2\right)
     \right]} {||\vec{r}_1-\vec{r}_2||}\\*
   &= \left(\frac{2\as}{\pi}\right)^2\left(\frac{\aB}{r_*}\right)^2\frac{e^2}{r_*}\ .
  \end{split}
\end{equation}

Here it is reasonable to assume that the optimal value of the variational parameter~$r_*$
will be found of the order of magnitude of the Bohr radius~$\aB$; and this implies that
the magnetic interaction energy~$\hmER$ is smaller than its electric counterpart~$\heER$
(\ref{eq:III.77}) by the factor~$\as^2\lesssim 10^{-4}$. This is also the order of
magnitude of the other relativistic effects; and therefore the groundstate energy
difference of ortho- and para-positronium cannot be expected to be properly predicted by
the present purely magnetic result (\ref{eq:IV.59}) (see below).
%%% Local Variables: 
%%% mode: latex
%%% TeX-master: "main_minenergy"
%%% End: 

\section{Positronium Groundstate}
\indent

For the situation where both particles masses are identical $(\Me=\Mp\doteqdot M)$, it is
reasonable to assume that both the first (positively charged) particle~$(a=1)$ and the
second (negatively charged) particle~$(a=2)$ do always occupy physically equivalent states.
According to this assumption, the positronium energy spectrum is expected to be
essentially a one-particle spectrum which is in perfect agreement with the observational
data~\cite{LeWe}. The conventional classification of the positronium energy levels relies on
the composition law for angular momenta so that the groundstate appears as the
doublet~$1{}^1S_0$ and~$1{}^3S_1$ corresponding to whether the total spin~$S$ is zero
($S=s_1-s_2=0$; \emph{para-positronium}) or is unity ($S=s_1+s_2=1$;
\emph{ortho-positronium}), see e.g. ref.~\cite{GrRe}. However in RST as a fluid-dynamical
theory, it is more adequate to base the classification upon the relative orientation of
both magnetic fields~$\vec{H}_a(\vec{r})$ rather than upon the angular momentum
composition law which is adequate for the conventional tensor product of Hilbert spaces
but not for the present Whitney sum of single-particle bundles.

But in any case, the inclusion of the magnetic (i.e.\ spin-spin) interactions is an
additional complication; and it is therefore convenient to first simplify the problem by
neglecting the magnetic interactions completely and considering the residual problem alone
($\leadsto$~\emph{``electrostatic approximation''}). In the non-relativistic conventional
theory, this truncated problem is then described by the two-particle Hamiltonian~$\hHS$
(\ref{eq:III.34}) and can be solved exactly by introducing the relative and center-of-mass
coordinates~\cite{Me}. The corresponding conventional groundstate
energy~$E_0|_{\mathrm{con}}$ is then easily found as
\begin{equation}
  \label{eq:V.1}
  E_0|_{\mathrm{con}}=-\frac{1}{4}\frac{e^2}{\aB}=-\frac{1}{4}\as^2\cdot Mc^2\simeq-6,80\, [eV]
\end{equation}
where~$\aB(=\hbar^2/Me^2)$ is the Bohr radius and~$\as(=e^2/\hbar c)$ is the fine
structure constant.

Indeed, this result~(\ref{eq:V.1}) is nothing else than the conventional hydrogen
groundstate energy due to a fixed nucleus, with merely the electron mass~$M$ being
replaced by the reduced mass~$M/2$ due to the comoving positron. Naturally, one will
demand from any new theory of quantum matter that it should reproduce this standard
result~(\ref{eq:V.1}) in its lowest order of approximation; and afterwards one may proceed
to compare the higher-order predictions of the various theoretical approaches. Therefore
we will now first clarify the way in which the standard result~(\ref{eq:V.1}) emerges in
RST, and afterwards one can turn to the magnetic effects as small corrections of the
electrostatic results. Amazingly enough, we will recover just the standard
result~(\ref{eq:V.1}) as an approximate (i.e.\ variational) solution within the RST
framework, namely by resorting to the RST principle of minimal
energy~$\delta\tEnT=0$~(\ref{eq:IV.26}).

\begin{center}
  \emph{\textbf{A.\ Electrostatic Approximation}}
\end{center}

Reasonably, the \emph{electric} properties of both particles may be adopted to be
approximately independent of the different \emph{magnetic} arrangements. Thus the Dirac
densities~$\akn$ can be assumed to be the same for both particles:
\begin{equation}
  \label{eq:V.2}
  \ikn\equiv\iikn\doteqdot\bkn\ .
\end{equation}
Furthermore, since these charge densities generate the electric potentials $\aAn$
according to the Poisson equations~(\ref{eq:III.47a})-(\ref{eq:III.47b}), both potentials
can differ at most in sign, i.e.
\begin{equation}
  \label{eq:V.3}
  \iAn\equiv-\iiAn\doteqdot\pAn\ ,
\end{equation}
with the common potential~$\pAn$ obeying the Poisson equation
\begin{equation}
  \label{eq:V.4}
  \Delta\pAn = -4\pi\as\bkn\ .
\end{equation}
If we resort here to the non-relativistic approximations
(\ref{eq:IV.23a})-(\ref{eq:IV.23b}) and \emph{tentatively} put for the non-relativistic wave amplitudes,
cf.~(\ref{eq:III.68}),
\begin{equation}
  \label{eq:V.5}
  \iiSp\equiv\iRp\Rightarrow\tilde{R}(r)=\sqrt{\btknn}\doteqdot\sqrt{\frac{8}{\pi r_*^2}}
  \cdot\exp\left(-\frac{r}{r_*} \right)
\end{equation}
then we just recover the former model potential~$\ppAn$~(\ref{eq:III.73}) as the common
potential~(\ref{eq:V.3}) for both particles. And correspondingly, their electrostatic
interaction energy~$\heER$~(\ref{eq:IV.22a}) is then just given by
equation~(\ref{eq:III.77}). 

Here it is important to remark that, due to the non-relativistic trial
function~$\tilde{R}(r)$~(\ref{eq:V.5}), the potential~$\ppAn$ is the \emph{exact}
solution of the non-relativistic Poisson equation~(\ref{eq:III.72}) and therefore the
Poisson constraints (\ref{eq:III.58})-(\ref{eq:III.59a}) are \emph{exactly} satisfied in
their non-relativistic form~(\ref{eq:IV.24a})-(\ref{eq:IV.24b}). Furthermore, it is easy
to see that the non-relativistic trial function~$\tilde{R}(r)$~(\ref{eq:V.5}) actually
obeys the normalization conditions~(\ref{eq:IV.25a})-(\ref{eq:IV.25b}). Thus both
constraints for the (second line of the) non-relativistic energy
functional~$\tEnT$ (\ref{eq:IV.26}) are automatically satisfied by our spherically
symmetric trial wave amplitude~$\tilde{R}(r)$~(\ref{eq:V.5}); and therefore one is
concerned solely with the physical contributions (first line) to the energy functional.

However, since the field energy~$\heER$ (as the interaction energy of both particles) is
already specified by equation~(\ref{eq:III.77}), one is left with the determination of the
non-relativistic kinetic energies~$\Ekin_{(a)}$~(\ref{eq:IV.20a})-(\ref{eq:IV.20b}).
Observing here the spherical symmetry of our trial function~(\ref{eq:V.5}) together with
the fact that both kinetic energies must be identical (i.e.~$\Ekin_{(1)}=\Ekin_{(2)}$),
one arrives at the total kinetic energy~$\Ekin$ as
\begin{equation}
  \label{eq:V.6}
  \Ekin\doteqdot\Ekin_{(1)}+\Ekin_{(2)}=\frac{\hbar^2}{2M}\int d^2\vec{r}\;
  \left(\frac{d\tilde{R}(r)}{dr} \right)^2 = \frac{\hbar^2}{Mr_*^2}\ .
\end{equation}
Consequently, the value of the non-relativistic functional~$\tEnT$ upon our spherically
symmetric trial function~$\tilde{R}(r)$~(\ref{eq:V.5}) becomes the following ordinary
function~$\tEnT(r_*)$ of the ansatz parameter~$r_*$:
\begin{equation}
  \label{eq:V.7}
  \tEnT(r_*)=\Ekin(r_*)+\heER(r_*)=\frac{\hbar^2}{Mr_*^2}-\frac{e^2}{r_*}\ .
\end{equation}

According to the established \emph{principle of minimal energy}, the positronium
groundstate energy~$E_0$ in the spherically symmetric approximation is given by the
minimal value of this function~$\tEnT(r_*)$~(\ref{eq:V.7}), i.e.
\begin{equation}
  \label{eq:V.8}
  E_0 = \tEnT\bigg|_\mathrm{min} = - \frac{e^2}{4\aB}\ ,
\end{equation}
and this minimum occurs for the value~$r_*\Rightarrow 2\aB$ of the ansatz parameter~$r_*$.
Thus the non-relativistic approximation of the RST principle of minimal energy yields just
the conventional Schr\"odinger value~(\ref{eq:V.1}) for the positronium groundstate!  This,
however, is surely an amazing result in a two-fold way:
\begin{itemize}
\item[\textbf{(i)}]
  Despite the very different mathematical structure of both approaches
  (Whitney sum vs.\ tensor product) the present RST prediction~(\ref{eq:V.8})
  coincides \emph{exactly} with the conventional Schr\"odinger prediction (\ref{eq:V.1}).
\item[\textbf{(ii)}] However, in contrast to the conventional prediction, which is adopted
  to be \emph{exact} within the standard framework of quantum mechanics, the corresponding
  RST prediction~(\ref{eq:V.8}) is based upon the choice of an appropriate trial function,
  cf.~(\ref{eq:V.5}), and therefore is an \emph{approximate} result within the RST
  framework. Thus the interesting question arises how close the \emph{exact} (but
  non-relativistic) RST prediction would come to the exact conventional
  prediction~(\ref{eq:V.1})?  

  Surely this is a difficult question because its answer would require to find the exact
  solution of the non-relativistic RST eigenvalue problem which consists of the
  (non-relativistic) eigenvalue equations~(\ref{eq:IV.18a})-(\ref{eq:IV.18b}) and the
  coupled (non-relativistic) Poisson equations~(\ref{eq:IV.29a})-(\ref{eq:IV.29b}).
\end{itemize}

\begin{center}
  \emph{\textbf{B.\ Hyperfine Splitting}}
\end{center}

The effect of level splitting by the magnetic (i.e.\ spin-spin) interactions is
experimentally well established and is found to amount to~$0,0008\ldots [eV]$ for the
positronium groundstate~\cite{LeWe}. It should be clear that such a small energy difference
between the triplet~$(\trip)$ and singlet~$(\sing)$ state falls into the order of
magnitude of the relativistic effects which such compete with the magnetic interaction
effects. Therefore it seems very unlikely that the total energy difference due to the
hyperfine splitting of the groundstate should be caused by the magnetic effects alone, but
nevertheless it may be interesting to estimate their relative contribution to the
hyperfine splitting of the groundstate within the present framework of RST.

For this purpose, one first has to demonstrate the specific way in which this level
dichotomy does emerge in RST. Such an effect, however, is rather obvious since in the
non-relativistic limit one can alternatively put the spin-down component~$\iiSp$
(\ref{eq:III.62a}) to zero and retain only the spin-up component~$\iiRp$ which itself must
then coincide with the first wave amplitude~$\iRp$; i.e.\ in place of the former
arrangement (\ref{eq:IV.58a})-(\ref{eq:IV.58b}) one puts now for the non-vanishing wave
amplitudes
\begin{subequations}
  \begin{align}
    \label{eq:V.9a}
    \iiRp\equiv\iRp\doteqdot&\tilde{R}(r) = \sqrt{\frac{8}{\pi
        r_*^2}}\exp\left(-\frac{r}{r_*}\right)\\*
    \label{eq:V.9b}
    \iSp&=\iiSp\equiv 0\ .
  \end{align}
\end{subequations}
This yields the parallelity of both Dirac
currents~$\vec{k}_a(\vec{r})(={}^{(a)}k_\phi\vec{e}_\phi)$,
\begin{equation}
  \label{eq:V.10}
  {}^{(a)}k_\phi=\frac{\hbar}{2M c}\cdot\frac{\tilde{R}(r)\cdot\frac{\partial}{\partial
      r}\tilde{R}(r)}{2\pi r}\hspace{5mm} (a=1,2)\ ,
\end{equation}
in contrast to the antiparallelity of the former case (\ref{eq:IV.35a})-(\ref{eq:IV.35b}).
According to the relationships (\ref{eq:II.36a})-(\ref{eq:II.36b}) and
(\ref{eq:II.43a})-(\ref{eq:II.43b}) between the Dirac currents~$k_{a\mu}$ and Maxwell
currents~${j^a}_\mu$, the Maxwell equations (\ref{eq:II.27a})-(\ref{eq:II.27b}) in
three-vector notation (\ref{eq:IV.31}) say that the parallelity of the Dirac
currents~$\vec{k}_a$ imply the antiparallelity of the magnetic fields~$\vec{H}_a$ and vice
versa. These arrangements of the RST fields suggest the following magnetic classification
of the positronium states~\cite{BMS}

\parbox[c]{4cm}{\textbf{ortho-positronium}\\ \centering$({}^1\!S_0)$}
\parbox[c]{9cm}{
\begin{subequations}
  \begin{align}
    \label{eq:V.11a}
    \vec{k}_1 &\equiv -\vec{k}_2 \doteqdot \vec{k}_{\mathrm{p}} = {}^{(p)}k_\phi\vec{e}_\phi\\*
    \label{eq:V.11b}
    \vec{j}_1 &\equiv \vec{k}_1 =\vec{k}_{\mathrm{p}};\vec{j}_2 \equiv -\vec{k}_2 = \vec{k}_{\mathrm{p}}\\*
    \label{eq:V.11c}
    \vec{A}_1 &\equiv \vec{A}_2 \doteqdot \vec{A}_{\mathrm{b}} = {}^{(b)}A_\phi\vec{e}_\phi\\*
    \label{eq:V.11d}
    \vec{H}_1 &\equiv \vec{H}_2 \doteqdot \vec{H}_{\mathrm{b}} =
    {}^{(b)}H_r\vec{e}_r+{}^{(b)}\vec{H}_\vartheta\vec{e}_\vartheta\ ,
  \end{align}
\end{subequations}
}

and analogously

\parbox[c]{4cm}{\textbf{para-positronium}\\ \centering$({}^3\!S_1)$}
\parbox[c]{9cm}{
\begin{subequations}
  \begin{align}
    \label{eq:V.12a}
    \vec{k}_1 &\equiv \vec{k}_2 \doteqdot \vec{k}_{\mathrm{b}} = {}^{(b)}k_\phi\vec{e}_\phi\\*
    \label{eq:V.12b}
    \vec{j}_1 &\equiv \vec{k}_1 =\vec{k}_{\mathrm{b}};\vec{j}_2 \equiv -\vec{k}_2 =-\vec{k}_{\mathrm{b}}\\*
    \label{eq:V.12c}
    \vec{A}_1 &\equiv -\vec{A}_2 \doteqdot \vec{A}_{\mathrm{p}} = {}^{(p)}A_\phi\vec{e}_\phi\\*
    \label{eq:V.12d}
    \vec{H}_1 &\equiv -\vec{H}_2 \doteqdot \vec{H}_{\mathrm{p}} =
    {}^{(p)}H_r\vec{e}_r+{}^{(p)}\vec{H}_\vartheta\vec{e}_\vartheta\ .
  \end{align}
\end{subequations}
}
Thus the present RST ortho-positronium corresponds to the conventional singlet
states~$(\sing)$ and RST para-positronium to the triplet states~$(\trip)$. It is true, this
RST classification of the positronium states is based upon the (\emph{non-relativistic})
decoupling of the spin-up and spin-down configurations~\cite{BMS} but is assumed to hold
also for the relativistic case where the spin-up and spin-down amplitudes~$\aRpm,\aSpm$
remain coupled so that spherically symmetric configurations are not possible, see the
eigenvalue equations (\ref{eq:III.66a})-(\ref{eq:III.66d}).

The present magnetic dichotomy of the positronium states lends itself now to a very
simple calculation of the groundstate hyperfine splitting. Namely, for a lowest-order
estimate one may resort to the two trial configurations of the parallel magnetic-fields
(\ref{eq:IV.58a})-(\ref{eq:IV.58b}) ($\leadsto$ ortho-positronium)  or of antiparallel
fields (\ref{eq:V.9a})-(\ref{eq:V.9b}) ($\leadsto$ para-positronium). The corresponding
magnetic field energy~$\hmER$ (\ref{eq:IV.1b}) differs then only in sign:
\begin{subequations}
  \begin{align}
  \label{eq:V.13a}
  \hmER|_\mathrm{ortho} &= \frac{\hbar c}{4\pi\as}\int d^3\vec{r}\;||\vec{H}_\mathrm{b}
  ||^2\equiv -\hmER|_\mathrm{para}\\*
  \label{eq:V.13b}
    \hmER|_\mathrm{para} &= -\frac{\hbar c}{4\pi\as}\int d^3\vec{r}\;||\vec{H}_\mathrm{p}
    ||^2\equiv-\hmER|_\mathrm{ortho}\ ,
  \end{align}
\end{subequations}
provided the magnetic field~$\vec{H}_{\mathrm{b}/\mathrm{p}}$ is computed approximately by
means of the trial functions for the amplitude combinations~$\{\iRp,\iiSp\}$
and~$\{\iRp,\iiRp\}$, as demonstrated in subsection IV.C (\emph{Magnetic Interactions}).
Thus referring to the magnetic interaction energy~$\hmER$ (\ref{eq:IV.59}) of
ortho-positronium, one ends up with the following total energy~$\tEnT$:
\begin{equation}
  \label{eq:V.14}
  \tEnT(r_*) = \frac{\hbar^2}{M r_*^2} - \frac{e^2}{r_*} \mp
  \left(\frac{2\as\aB}{\pi}\right)^2\cdot\frac{e^2}{r_*^3}\ ,
\end{equation}
which of course is the magnetic generalization of the simpler purely electric case
(\ref{eq:V.7}). (The upper/lower sign refers to the ortho/para-configurations
(\ref{eq:V.11a})-(\ref{eq:V.12d})).

The minimal value of the total energy~$\tEnT(r_*)$ occurs now at the slightly shifted
position
\begin{equation}
  \label{eq:V.15}
  r_\mathrm{min} = \aB\left[1+\sqrt{1\mp3\left(\frac{2\as}{\pi}\right)^2}\right]
 \simeq 2\aB\left[1\mp 3\left(\frac{\as}{\pi}\right)^2 \right]\ ,
\end{equation}
and the corresponding minimal value of the energy becomes now in the order of~$\as^2$
\begin{equation}
  \label{eq:V.16}
  \tEnTmin = -\frac{e^2}{4\aB}\left[1\pm 2\left(\frac{\as}{\pi}\right)^2 \right]\ .
\end{equation}
Therefore the hyperfine splitting~$\Delta\tEnT$ is predicted by the present estimate
as
\begin{equation}
  \label{eq:V.17}
  \Delta\tEnT \doteqdot \tEnT|_\mathrm{ortho} - \tEnT|_\mathrm{para} \equiv -2\hmER
  =  -\left(\frac{2\as}{\pi}\right)^2 \frac{e^2}{4\aB}\ .
\end{equation}
This is much smaller than the electrostatic binding energy of~$6,80\, [eV]$, cf.\
(\ref{eq:V.8}), namely
\begin{equation}
  \label{eq:V.18}
  \Delta \tEnT = -(2,15\cdot 10^{-5})\cdot 6,80\, [eV] = - 1,46\cdot 10^{-4}\, [eV]\ .
\end{equation}

It is true, this is qualitatively in agreement with the experimental fact that the binding
energy of the ortho-system is greater than that of the para-system~\cite{LeWe}; but the
experimental value of the hyperfine splitting is~$-8,41\cdot 10^{-4}\, [eV]$ which is six
times larger than the present RST prediction (\ref{eq:V.18}). Such a discrepancy may be
understood in the sense that for the positronium hyperfine splitting in the order
of~$\as^2$ it is necessary to use some trial function~$\tilde{R}(r)$ which is closer to
the exact solution than the simple exponential function (\ref{eq:V.9a}). In any case, it
seems worthwhile to look for the exact solutions of both the relativistic and
non-relativistic RST eigenvalue problem in order to test its theoretical accuracy in
comparison to the experimental situation and the other theoretical approaches.
%%% Local Variables: 
%%% mode: latex
%%% TeX-master: "main_minenergy"
%%% End: 

%%% Local Variables: 
%%% mode: latex
%%% TeX-master: "main_minenergy"
%%% End: 


\begin{thebibliography}{99}
\bibitem{We} S.\ Weinberg, \emph{The Quantum Theory of Fields}, vol.~1-3, Cambridge
  University Press (1996)

\bibitem{Gr} F.\ Gross, \emph{Relativistic Quantum Mechanics and Field Theory}, John Wiley
  \& Sons, New York (1999)

\bibitem{La} C.\ Lanczos, \emph{The Variational Principles of Mechanics}, Dover (1986)

\bibitem{Ru} H.\ Rund, \emph{The Hamiltonian-Jacobi Theory in the Calculus of Variations},
  van Norstrand (1966)

\bibitem{BeSo} T.\ Beck and M.\ Sorg, \emph{Positive and Negative Charges in Relativistic
    Schr\"odinger Theory}, http://arxiv.org/abs/hep-th/0609164

\bibitem{BMS} T.\ Beck, M.\ Mattes and M.\ Sorg, \emph{Positronium Groundstate in
    Relativistic Schr\"odinger Theory}, http://arxiv.org/abs/0704.3810

\bibitem{BeSo2} T.\ Beck and M.\ Sorg, \emph{Two- and Three-Particle Systems in Relativistic
    Schr\"odinger Theory}, Found.\ Phys.\ \textbf{37}, 1093 (2007)

\bibitem{LeWe} R.\ Ley and G.\ Werth, \emph{Lecture Notes in Physics} \textbf{570},
 ed.\ by S.\ G.\ Karshenboim et al., p.~407-418, Springer (2001)

\bibitem{SSMS} P.\ Schust, F.\ Stary, M.\ Mattes and M.\ Sorg, Found.\ Phys.\ \textbf{35}, 1043 (2005)

\bibitem{PS} S.\ Pruss-Hunzinger and M.\ Sorg, Nuov.\ Cim.\ \textbf{118 B}, 903 (2003)

\bibitem{BeSa} H.\ A.\ Bethe and E.\ E.\ Salpeter, \emph{Quantum Mechanics of One- and
    Two-Electron Atoms}, Springer (1957)

\bibitem{Fl} S.\ Fl\"ugge, \emph{Practical Quantum Mechanics}, Springer (1974)

\bibitem{TDL} C.\ Cohen-Tannoudji, B.\ Diu and F.\ Lalo\"e, \emph{Quantum Mechanics},
  vol.~II, John Wiley (1977)

\bibitem{Ba} L.\ E.\ Ballentine, \emph{Quantum Mechanics}, World Scientific (1999)

\bibitem{Wi} S.\ Wilson (ed.), \emph{Methods in Computational Chemistry}, vol~2:
  \emph{Relativistic Effects in Atoms and Molecules}, Plenum Press (1988)

\bibitem{GrRe} W.\ Greiner and J.\ Reinhardt, \emph{Field Quantization}, Springer (1996)

\bibitem{Me} A.\ Messiah, \emph{Quantum Mechanics}, vol.~I, North-Holland (1965)


\end{thebibliography}
\end{document}